\documentclass[twoside,twocolumn,9pt]{article}
\usepackage{extsizes}
\usepackage[super,sort&compress,comma]{natbib} 
\usepackage[version=3]{mhchem}
\usepackage[left=1.5cm, right=1.5cm, top=1.785cm, bottom=2.0cm]{geometry}
\usepackage{balance}
\usepackage{mathptmx}
\usepackage{sectsty}
\usepackage{graphicx} 
\usepackage{lastpage}
\usepackage[format=plain,justification=justified,singlelinecheck=false,font={stretch=1.125,small,sf},labelfont=bf,labelsep=space]{caption}
\usepackage{float}
\usepackage{fancyhdr}
\usepackage{fnpos}
\usepackage[english]{babel}
\addto{\captionsenglish}{%
  
}
\usepackage{array}
\usepackage{droidsans}
\usepackage{charter}
\usepackage[T1]{fontenc}
\usepackage[usenames,dvipsnames]{xcolor}
\usepackage{setspace}
\usepackage[compact]{titlesec}
\usepackage{hyperref}
\usepackage{comment}
\usepackage{chemformula}
\newcommand{\orcid}[1]{\href{https://orcid.org/#1}{\hskip2pt\includegraphics[width=9pt]{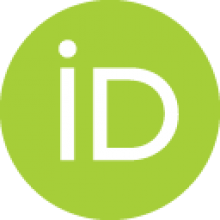}}}

\definecolor{cream}{RGB}{222,217,201}

\begin{document}

\pagestyle{fancy}
\thispagestyle{plain}
\fancypagestyle{plain}{
\renewcommand{\headrulewidth}{0pt}
}

\makeFNbottom
\makeatletter
\renewcommand\LARGE{\@setfontsize\LARGE{15pt}{17}}
\renewcommand\Large{\@setfontsize\Large{12pt}{14}}
\renewcommand\large{\@setfontsize\large{10pt}{12}}
\renewcommand\footnotesize{\@setfontsize\footnotesize{7pt}{10}}
\makeatother

\renewcommand{\thefootnote}{\fnsymbol{footnote}}
\renewcommand\footnoterule{\vspace*{1pt}%
\color{cream}\hrule width 3.5in height 0.4pt \color{black}\vspace*{5pt}} 
\setcounter{secnumdepth}{5}

\makeatletter 
\renewcommand\@biblabel[1]{#1}            
\renewcommand\@makefntext[1]%
{\noindent\makebox[0pt][r]{\@thefnmark\,}#1}
\makeatother 
\renewcommand{\figurename}{\small{Fig.}~}
\sectionfont{\sffamily\Large}
\subsectionfont{\normalsize}
\subsubsectionfont{\bf}
\setstretch{1.125} 
\setlength{\skip\footins}{0.8cm}
\setlength{\footnotesep}{0.25cm}
\setlength{\jot}{10pt}
\titlespacing*{\section}{0pt}{4pt}{4pt}
\titlespacing*{\subsection}{0pt}{15pt}{1pt}

\fancyfoot{}
\fancyfoot[LO,RE]{\vspace{-7.1pt}\includegraphics[height=9pt]{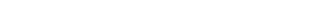}}
\fancyfoot[CO]{\vspace{-7.1pt}\hspace{11.9cm}\includegraphics{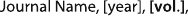}}
\fancyfoot[CE]{\vspace{-7.2pt}\hspace{-13.2cm}\includegraphics{head_foot/RF}}
\fancyfoot[RO]{\footnotesize{\sffamily{1--\pageref{LastPage} ~\textbar  \hspace{2pt}\thepage}}}
\fancyfoot[LE]{\footnotesize{\sffamily{\thepage~\textbar\hspace{4.65cm} 1--\pageref{LastPage}}}}
\fancyhead{}
\renewcommand{\headrulewidth}{0pt} 
\renewcommand{\footrulewidth}{0pt}
\setlength{\arrayrulewidth}{1pt}
\setlength{\columnsep}{6.5mm}
\setlength\bibsep{1pt}

\makeatletter 
\newlength{\figrulesep} 
\setlength{\figrulesep}{0.5\textfloatsep} 

\newcommand{\topfigrule}{\vspace*{-1pt}%
\noindent{\color{cream}\rule[-\figrulesep]{\columnwidth}{1.5pt}} }

\newcommand{\botfigrule}{\vspace*{-2pt}%
\noindent{\color{cream}\rule[\figrulesep]{\columnwidth}{1.5pt}} }

\newcommand{\dblfigrule}{\vspace*{-1pt}%
\noindent{\color{cream}\rule[-\figrulesep]{\textwidth}{1.5pt}} }

\makeatother

\twocolumn[
  \begin{@twocolumnfalse}
{\includegraphics[height=30pt]{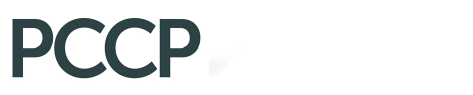}\hfill\raisebox{0pt}[0pt][0pt]{\includegraphics[height=55pt]{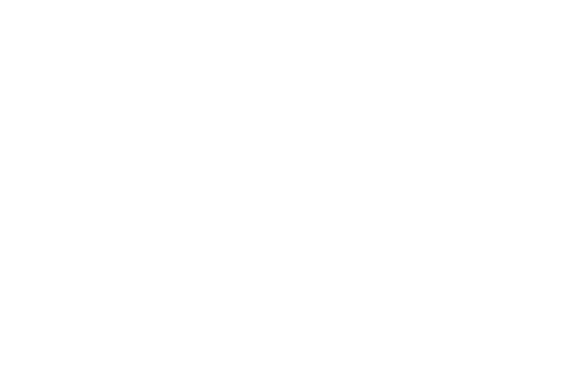}}\\[1ex]
\includegraphics[width=18.5cm]{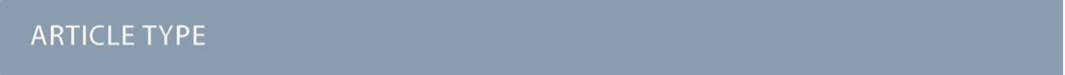}}\par
\vspace{1em}
\sffamily
\begin{tabular}{m{4.5cm} p{13.5cm} }

\includegraphics{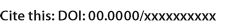} & \noindent\LARGE{\textbf{Binding energies of ethanol and ethylamine on interstellar water ices: synergy between theory and experiments$^\dag$}} \\
\vspace{0.3cm} & \vspace{0.3cm} \\

 & \noindent\large{Jessica Perrero,\orcid{0000-0003-2161-9120} \textit{$^{a,b}$} Julie Vitorino,\orcid{0000-0003-2502-487X} \textit{$^{c}$} Emanuele Congiu,\orcid{0000-0003-0313-826X} \textit{$^{c}$} Piero Ugliengo,\orcid{0000-0001-8886-9832} $^{\ast}$\textit{$^{b}$} Albert Rimola,\orcid{0000-0002-9637-4554} $^{\ast}$\textit{$^{a}$} and François Dulieu\orcid{0000-0001-6981-0421} $^{\ast}$\textit{$^{c}$}} \\

\includegraphics{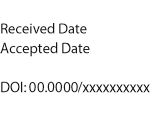} & \noindent\normalsize{Experimental and computational chemistry are two disciplines to conduct research in Astrochemistry, providing essential reference data for both astronomical observations and modeling. These approaches not only mutually support each other, but also serve as complementary tools to overcome their respective limitations. Leveraging on such synergy, we characterized the binding energies (BEs) of ethanol (\ch{CH3CH2OH}) and ethylamine (\ch{CH3CH2NH2}), two interstellar complex organic molecules (iCOMs), onto crystalline and amorphous water ices through density functional theory (DFT) calculations and temperature programmed desorption (TPD) experiments. Experimentally, \ch{CH3CH2OH} and \ch{CH3CH2NH2} behave similarly, 
in which desorption temperatures are higher on the water ices than on a bare gold surface. Computed cohesive energies of pure ethanol and ethylamine bulk structures allow describing the BEs of the pure species deposited on the gold surface, as extracted from the TPD curve analyses. The BEs of submonolayer coverages of \ch{CH3CH2OH} and \ch{CH3CH2NH2} on the water ices cannot be directly extracted from TPD due to their co-desorption with water, but they are computed through DFT calculations, and found to be greater than the cohesive energy of water.
The behaviour of \ch{CH3CH2OH} and \ch{CH3CH2NH2} is different when depositing adsorbate multilayers on the amorphous ice, in that, according to their computed cohesive energies, ethylamine layers present weaker interactions compared to ethanol and water. Finally, from the computed BEs of ethanol, ethylamine and water, we can infer that the snow-lines of these three species in protoplanetary disks will be situated at different distances from the central star. It appears that a fraction of ethanol and ethylamine is already frozen on the grains in the water snow-lines, causing their incorporation in water-rich planetesimals.
} 

\end{tabular}

 \end{@twocolumnfalse} \vspace{0.6cm}

  ]


\renewcommand*\rmdefault{bch}\normalfont\upshape
\rmfamily
\section*{}
\vspace{-1cm}


\footnotetext{\textit{$^{a}$~Departament de Qu\'{i}mica, Universitat Aut\`{o}noma de Barcelona, Bellaterra, 08193, Catalonia, Spain. E-mail: albert.rimola@uab.cat}}
\footnotetext{\textit{$^{b}$~Dipartimento di Chimica and Nanostructured Interfaces and Surfaces (NIS) Centre, Universit\`{a} degli Studi di Torino, via P. Giuria 7, 10125, Torino, Italy. E-mail: piero.ugliengo@unito.it}}
\footnotetext{\textit{$^{c}$~CY Cergy Paris Université, Observatoire de Paris, PSL University, Sorbonne Université, CNRS, LERMA, F-95000 Cergy, France. E-mail: francois.dulieu@cyu.fr}}

\footnotetext{\dag~Electronic Supplementary Information (ESI) available at DOI: \url{https://doi.org/xxxxxxxx}. Computational models available at} \url{https://zenodo.org/10.5281/zenodo.11103454}



\section{Introduction} \label{sec:intro}

In astrochemistry, ethanol (\ch{CH3CH2OH}) and ethylamine (\ch{CH3CH2NH2}) belong to a class of compounds called interstellar complex organic molecules (iCOMs). These are compunds that contain between 6--12 atoms, in which at least one is carbon, also bearing non-metal heteroatoms like, N, O, S or P.\cite{Herbst2009,Herbst2017,Ceccarelli_2017} These molecules are the simplest organic compounds synthesised in space, and hence they are referred to be the dawn of the organic chemistry. Additionally, they are thought to be precursors of more complex organic molecules, which can be of biological relevance, such as amino acids, nucleobases and sugars.\cite{caselli2012}

Ethylamine is thought to be the precursor of the alanine amino acid, just as methylamine is supposed to be the parent of glycine \cite{forstel2017,Hudson2022}. Methylamine and ethylamine have been observed in the coma of comet 67P/C-G together with glycine.\cite{goesmann2015} Furthermore, they have also been detected, together with a handful of amino acids, in comet 81P/Wild2 by Stardust mission.\cite{glavin2008} In the interstellar medium (ISM), ethylamine has only been tentatively detected towards the Galactic center cloud G+0.693-0.027.\cite{zeng2021} On the other hand, ethanol has been detected in several comets (e.g., 67P/C-G,\cite{Altwegg2019} Lovejoy,\cite{biver2015lovejoy} and Hale-Bopp\cite{crovisier2004}) and in both warm (e.g., Sgr B2,\cite{zuckerman1975} Orion,\cite{ohishi1988,pearson1997} and SVS-13\cite{bianchi2019}) and cold (e.g., L483\cite{agundez2019}, and TMC-1\cite{agundez2023}) environments of the ISM. Recently, it has also been identified in icy mantles.\cite{mcclure2023, rocha2024}


The presence of a complex molecule such as \ch{CH3CH2OH} in warm sources was surprising, but plausible due to the environmental temperatures. However, its detection in cold regions challenged astrochemists. The interest on ethanol grew after a correlation between the abundances of glycolaldehyde and ethanol in the protostellar shock region L1157-B1 was evidenced.\cite{lefloch2017} Consequently, a series of gas-phase reactions starting from ethanol and leading to glycolaldehyde were proposed, beginning with a H-abstraction step operated by Cl or OH radicals. From this neutral-neutral gas-phase mechanism, it appeared that ethanol could be the parent molecule of formic acid (HCOOH) and other iCOMs such as glycolaldehyde (\ch{HCOCH2OH}), acetic acid (\ch{CH3COOH}), and acetaldehyde (\ch{CH3CHO}).\cite{skouteris2018}

The reactivity of \ch{CH3CH2NH2}, due to its scarce detection in the ISM, has been poorly explored. The UV irradiation and thermal processing of methylamine-containing ices yielded a variety of products, such as formamide, ethylamine, methylcyanide and N-heterocycles, showing how molecular complexity can increase upon methylamine processing.\cite{carrascosa2021} 
The production of \ch{CH3CH2NH2} is usually attributed to the reaction between ammonia and C-bearing species on the ice mantle. The exposure of NH$_3$ and CH$_4$/C$_2$H$_6$ ices to cosmic ray radiation at 5 K resulted in the formation of CH$_3$NH$_2$ and \ch{CH3CH2NH2}, respectively.\cite{forstel2017} At higher temperatures, photolysis of C$_2$H$_2$ + NH$_3$ between 180 and 300 K produced ethylamine (among other products), with a yield proportional to temperature.\cite{keane2017}

Formation of ethanol, in contrast, has been more explored. Experimentally, both energetic and non-energetic pathways have been identified: UV irradiation of CO$_2$-rich and H$_2$O-rich ices;\cite{Hudson_2013} UV irradiation and radiolysis of \ch{C2H2}:\ch{H2O} ices;\cite{wu2002,chuang2021} O addition to ethane; \cite{bergner2019} and exposure of \ch{C2H2}:\ch{O2} ice to H atoms.\cite{chuang2020} Computational works also characterized ethanol formation, proposing synthetic routes on dust grains at low temperatures and through almost barrierless pathways. The radical-radical coupling between \ch{CH3} and \ch{CH2OH} on water ice was postulated as a promising mechanism, since its competitive channel (an H-abstraction yielding \ch{CH4} and \ch{H2CO}) was found to be highly disfavoured.\cite{enrique-romero2022} The reaction of CCH + \ch{H2O} (where water is a component of the ice mantle), passing through the formation of vinyl alcohol, can yield ethanol after some hydrogenation steps that, in case of presenting a barrier, can proceed via tunnelling.\cite{perrero2022ethanol} More recently, two works investigated the reactivity of atomic carbon on CO-rich and on H$_2$O-rich ices, suggesting that the two scenarios can lead to the formation of iCOMs in a non-energetic way, especially ethanol, when species characterized by a \ch{C\bond{double}C\bond{double}O} skeleton are fully hydrogenated.\cite{ferrero2023_c_co,ferrero2023_c_h2o}    
Clearly, the discrepancy in the quantity of studies regarding these two molecules is linked to their detection. Numerous factors may contribute to the elusive nature of ethylamine to astronomers. For instance, it might predominantly reside in ice mantles rather than in the gas phase; it could exhibit higher reactivity leading to rapid consumption upon entering the gas phase; or it may have spectral features that overlap with those of other molecules. To streamline our investigation, we focused on the first point. The adsorption and desorption processes are determined by the binding energy (BE), a pivotal parameter that quantifies the strength of the interaction between a molecule and a surface.\citep{minissale2022} It also regulates whether diffusion processes can take place, since diffusion barriers are usually assumed to be a fraction of the BE, \cite[e.g.,][]{karssemeijer2014,cuppen2017,kouchi2020,mate2020}.
The first estimates of \ch{CH3CH2OH} and \ch{CH3CH2NH2} BEs were made by \citet{garrod2013}, who proposed BE = 65.5 kJ mol$^{-1}$ for ethylamine and BE = 52.0 kJ mol$^{-1}$ for ethanol, in order to introduce them in astrochemical models.
In 2011, \citet{lattelais2011} determined the BE of \ch{CH3CH2OH} on water ice to be 56.5 kJ mol$^{-1}$.
More recently, \citet{etim2018} computed the BE of ethanol using the methodology introduced by \citet{wakelam2017}, that is, simulating the ice with one water molecule and calculating the BE (in Kelvin) using the scaling BE$_{ice}$ = 289.019 + (1.65174 $\times$ BE$_{1\ch{H2O}}$), this way obtaining a value between 37.4 kJ mol$^{-1}$ and 40.7 kJ mol$^{-1}$, with an uncertainty on the final result of 30\%. To the best of our knowledge, no other estimates are available.

For this reason, we characterized both experimentally and computationally the BEs of ethanol and ethylamine adsorbed on crystalline and amorphous water ice (CI and ASW, respectively). In order to interpret the results of the experiments, we also computed the bulk and surface cohesive energies of ethanol and ethylamine crystals, as well as the BEs of each species on the most stable surface formed during the growth of the corresponding crystal.

BEs, in addition to being a key parameter in astrochemical models, are also one of the two quantities needed to compare the theoretical calculations with the desorption energy E$_{\rm des}$ extracted from the experiments, together with the pre-exponential factor $\nu$.\cite{minissale2022} The latter is often defined as the vibrational frequency of a given species in its surface potential well, and it takes into account the entropic effects of the desorption process. The pre-exponential factor enters in the equation that determines the thermal desorption rate, together with the E$_{\rm des}$, which accounts for the enthalpic contribution. There are two main ways to determine $\nu$: one is to adopt the approximated approach of \citet{hasegawa1993}, the alternative consists in exploiting the transition state theory within the immobile adsorbate approximation, as \citet{tait2005} suggested. 
In this work, we adopted the latter approach to determine $\nu$, with the aim to derive the E$_{\rm des}$ from the experimental data. 

The work is organized as follows: in Section \ref{sec:theory} we present the computational results, while Section \ref{sec:experiments} is dedicated to the laboratory experiments. Discussion of the obtained results through the two approaches is provided in Section \ref{sec:discussion}, along with their astrophysical implications. Finally, we summarized the most important findings of the work in Section \ref{sec:conclusions}.

\section{Theoretical Simulations}\label{sec:theory}

For the sake of clarity, we adopted the following notation to distinguish between the molecular adsorbates and the solid-state phases: \ch{CH3CH2OH} and \ch{CH3CH2NH2} refers to the adsorbates, while the bulk and surfaces of ethanol and ethylamine are identified as EtOH and EtNH$_2$. Accordingly, the adsorption of ethanol on an ethanol surface is referred to as \ch{CH3CH2OH} on an EtOH surface.

\subsection{Methodology}\label{par:computational}

Quantum chemical periodic calculations were performed with the \textsc{CRYSTAL17} code, which is suitable to treat systems with zero (molecules), one (polymers), two (slab surfaces), and three (bulks) periodic dimensions. It uses localized Gaussian functions as basis sets, contrary to the plane waves adopted by most of the periodic codes.\cite{dovesi2018} 
This allows to rigorously define true periodic 2D models, avoiding to build replicas of the system along the non-periodic direction. In this work, we modelled 3D (bulks) and 2D (slabs) systems for \ch{H2O}, EtOH and EtNH$_2$ ices.

\subsubsection{Computational details}

All the geometry optimizations and frequency calculations were run with the semi-empirical HF-3c method, a Hartree-Fock-based method adopting a minimal basis set (MINI-1),\cite{tatewaki1980} to which three empirical corrections (3c) are added to make up for the smallness of the basis set:\cite{sure2013} (i) the Grimme’s D3 empirical term with the Becke-Johnson (BJ) damping scheme (D3(BJ)\cite{grimme2010,grimme2011}) to account for dispersion energy arising from for noncovalent interactions; (ii)
a short range bond correction to recover the systematically overestimated covalent bond lengths for electronegative elements;\cite{brandenburg2013} and (iii) the geometrical counterpoise (gCP) method developed by \citet{kruse2012} to \textit{a priori} remove the BSSE.

To compute the BEs, DFT single-point energy calculations were performed on each optimized HF-3c geometry (hereafter referred to as DFT//HF-3c). The hybrid B3LYP functional \cite{becke:1988,becke:1993,lyp:1988} combined with the D3(BJ) correction for the dispersion energy was used, a methodology already tested and applied for closed-shell species.\cite{ferrero2020, perrero2022} B3LYP-D3(BJ) was combined with an Ahlrichs triple zeta valence quality basis set supplemented with a double set of polarization functions, defined as A-VTZ* \cite{schafer:1992}.

All the structures (bulks, surfaces and adsorption complexes) were characterized by HF-3c harmonic frequency calculations performed on the entire system. This serves to confirm the nature of the stationary points and to compute the zero point energy (ZPE) correction, that accounts for the kinetic vibrational energy at 0 K.

\subsubsection{Water ice surface models and calculation of the BEs}

\begin{figure*}[ht]
\centering
  \includegraphics[width=\textwidth]{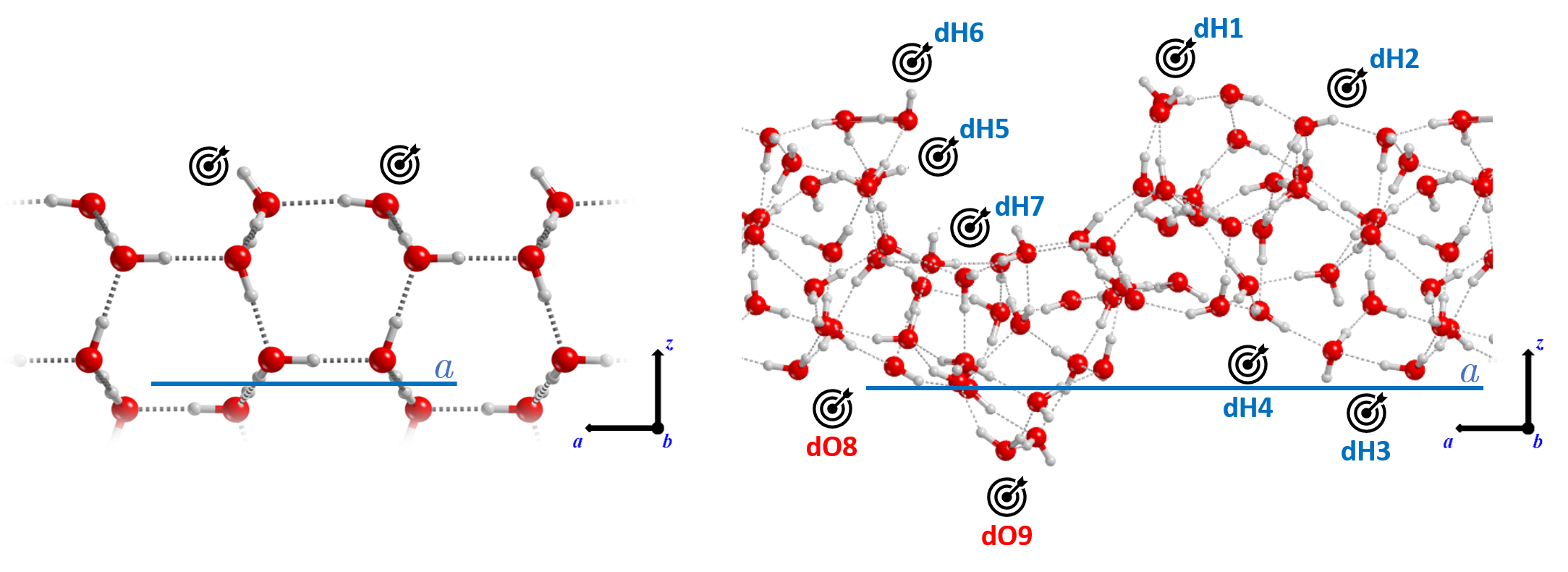} \\
  \caption{Crystalline (left) and amorphous (right) periodic water ice models adopted in this work. The target icon indicates the binding sites considered for the adsorption of \ch{CH3CH2OH} and \ch{CH3CH2NH2}. On the amorphous model the sites have been labeled as dangling hydrogen atoms (dH1--dH7) and dangling oxygen atoms (dO8--dO9).
  Colour code: red, oxygen; grey, hydrogen. The unit cell parameter is also highlighted (in blue).}
  \label{fig:ice_models}
\end{figure*}

Two periodic water ice models previously used in refs.\cite{ferrero2020, perrero2022} (see Figure \ref{fig:ice_models}) were used to compute the adsorption of ethanol and ethylamine. To simulate the surfaces, we adopted the slab model, in which a slab of a given thickness is cut out from the bulk ice model. The first structure is a crystalline ice model, represented by the (010) surface cut out from the bulk of the proton-ordered P-ice.\cite{casassa1997}
The unit cell is characterized by the cell parameters $\left| a \right|$ = 9.065 \r{A} and $\left| b \right|$ = 7.153 \r{A} and it consists of twelve atomic layers. However, interstellar ices are mostly amorphous in nature;\cite{boogert2015, mcclure2023} therefore, we adopted a more realistic model consisting of 60 disordered water molecules per unit cell. The structure possesses a cavity and several edges, thereby showing multiple different binding sites. Its cell parameters are: $\left| a \right|$ = 20.355 \r{A}, $\left| b \right|$ = 10.028 \r{A}, and $\left| \gamma \right|$ = 103.0°.

The adsorption complexes were manually constructed placing the adsorbates on the binding sites present in the ice models, following the principle of electrostatic complementarity between the adsorbate and the surface, and fully optimizing the structures. 
On the crystalline ice model, we characterized three situations: i) single molecule adsorption (S-ADS), represented by one molecule adsorbed per unit cell; ii) a half monolayer (H-ML) coverage, simulated by adsorbing two molecules per unit cell; 
and iii) a full monolayer (F-ML) coverage, modelled by adsorbing four molecules per unit cell.
Alternatively, on the amorphous ice model, a range of single adsorptions were characterized with the aim to obtain BEs describing the adsorbate/ice interactions of different strengths.
Nine positions suitable to adsorb \ch{CH3CH2OH} and \ch{CH3CH2NH2} were identified, among which seven are dangling H atoms (dH) and the remaining two are dangling O atoms (dO). Given the presence of both H-bond acceptor and donor groups in the adsorbates, such ensemble of binding sites allows the establishment of both types of interactions. 

The final binding energy, BE(0), defined as a positive quantity in case of favourable interaction, results from subtracting the basis set superposition error (BSSE), arising from the finiteness of the basis set, and adding the HF-3c ZPE correction at 0 Kelvin ($\Delta ZPE = ZPE_{complex} - ZPE_{ice} - ZPE_{species}$) to the bare interaction energy ($\Delta E = E_{complex} - E_{ice} - E_{species}$).
\begin{equation}
   BE(0) = BE - \Delta ZPE = - (\Delta E - BSSE) - \Delta ZPE
\end{equation}

Each BE value can be decomposed in a purely electronic contribution extracted from the pure DFT calculation and a fraction due to purely dispersive interactions, estimated with the D3(BJ) term. We are aware that this approach is not rigorous, thus our estimates should be taken as indicative. 
Detailed explanation of the computation of the BE(0)s and their energetic contributions can be found in the ESI$^\dag$ and in previously published works.\cite{ferrero2020,perrero2022}  

\subsubsection{Modeling of EtOH and EtNH$_2$ ices}


The EtOH and EtNH$_2$ ice bulk structures were modeled by fully optimizing their crystalline structures, determined by X-ray diffraction.\cite{Jonsson1976,maloney2014} Subsequently, several low Miller index surfaces were generated to characterize their stability. For each slab, the internal atomic positions were optimized while keeping frozen the unit cell parameters to the bulk values. The thickness of each slab was determined in such a way that their surface energy converged to a plateau.

The surface energy (E$_S$, in J/m$^{2}$) is the energy required to cut out a slab from the bulk. The most stable surfaces are characterized by the lowest surface energies. E$_S$ is calculated through the equation:
\begin{equation}
    E_S  = \frac{E_{slab} - N \cdot E_{bulk}}{2A}   
\end{equation}
where $E_{slab}$ is the energy of the surface, $E_{bulk}$ is the energy of the bulk, \textit{N} = $z_{slab}/z_{bulk}$ with \textit{z} being the number of molecules contained in the unit cells of slab and bulk, and \textit{A} is the surface area. The factor 2 accounts for the existence of two equal surfaces of the slab.

The cohesive energy (E$_C$, in kJ mol$^{-1}$) of the bulk (or a surface) results from the equation:
\begin{equation}
    E_C  = \frac{E_{bulk(slab)}}{z} - E_{mol}
\end{equation}
where $E_{mol}$ is the energy of the fully optimized isolated molecule in the gas phase.

The extraction energy (E$_X$) of a molecule from a surface is the energy cost to extract the molecule from the surface. It is obtained by calculating the energy of the structure that follows the removal of one surface molecule without re-optimization of the structure, and is repeated for each molecule located in the external layer of the surface:

\begin{equation}
    E_X = E_{carved} + E_{mol} - E_{slab} + \Delta ZPE_{extr}
\end{equation}
where E$_{carved}$ is the energy of the surface upon molecule extraction, and $\Delta ZPE_{extr}$ accounts for the contribution of the ZPE to the extraction processes, which is approximated to $\Delta ZPE_{extr} = ZPE_{surface}/N - ZPE_{mol}$.

The most stable surface of a crystal is the most extended one, and is characterized by the lowest E$_S$ and the highest E$_C$ values. In principle, it should also possess the highest E$_X$, since it is more difficult to extract a molecule from a well packed structure. The (010) \ch{EtOH} and the (100) \ch{EtNH2} surfaces satisfy these criteria. 
 
Therefore, we simulated the adsorption of \ch{CH3CH2OH} and \ch{CH3CH2NH2} onto these surfaces, identifying three binding sites per adsorbate.

\subsection{Results}
This section is organized as follows: subsection \ref{par:modelling} addresses the modeling of bulks and surfaces of EtOH and EtNH$_2$ crystals; subsection \ref{par:be_icom} regards the computation of the BE(0)s obtained on such surfaces; subsection \ref{par:be_ice} presents the BE(0)s for the S-ADS scenario on the crystalline and amorphous water ice surface models; and subsection \ref{par:be_monolayer} shows the results for the H-ML and F-ML coverages.

\subsubsection{EtOH and EtNH$_2$ crystal and surface ices}\label{par:modelling}

\begin{figure*}[htb]
\centering
  \includegraphics[width=\textwidth]{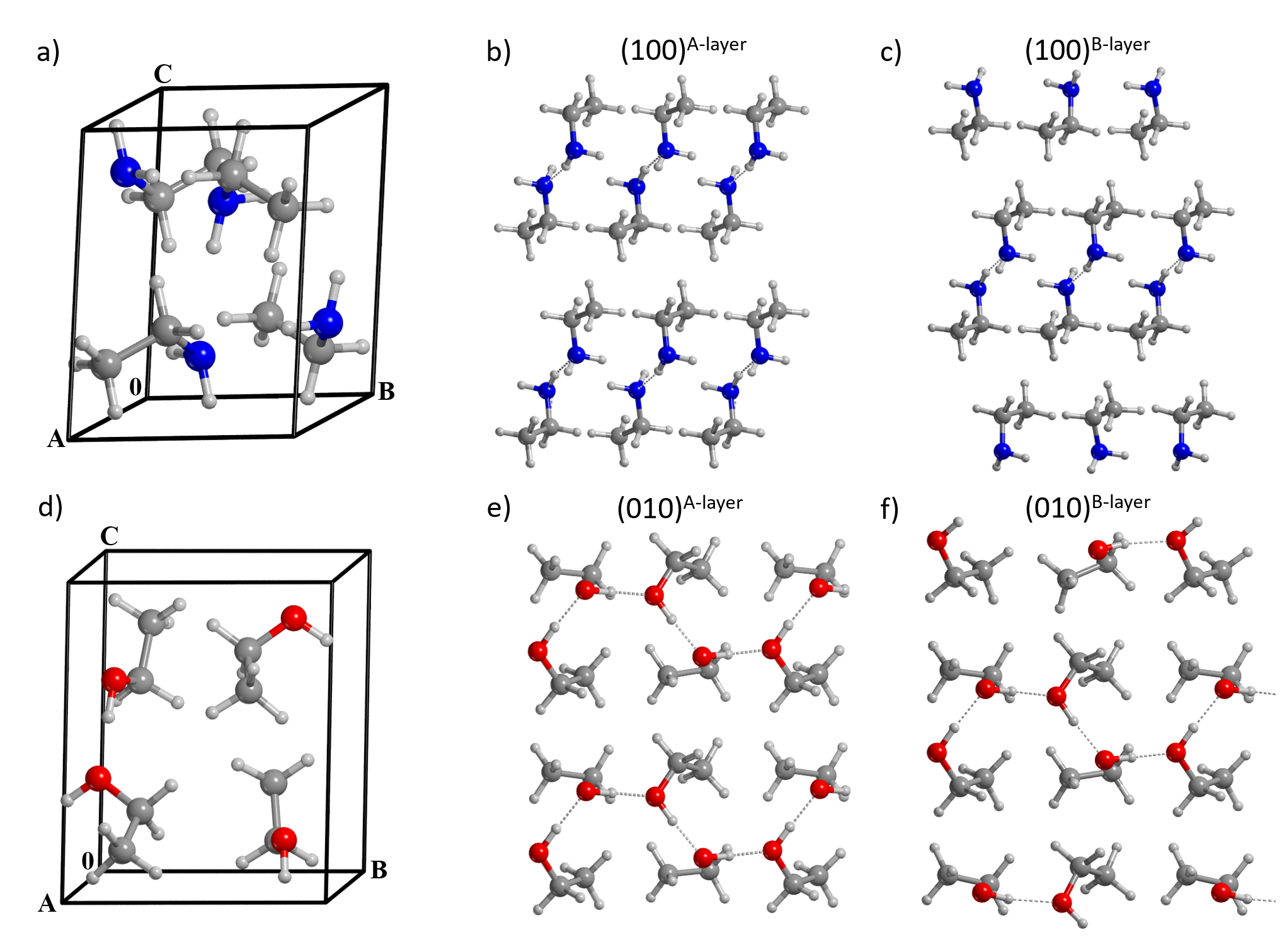} \\
  \caption{Unit cell of the bulk structure of \ch{EtNH2} (a) and EtOH (d). Lateral view of the (100) \ch{EtNH2} (b) and (010) EtOH (e) slabs, the most stable surfaces cut out from the respective bulks, exposing their apolar backbone to the external environment (A-layer type). Lateral view of the (100) \ch{EtNH2} (c) and (010) EtOH (f) slabs, which were generated by cutting out the surface in order to expose the polar moieties to the external environment (B-layer).}
  \label{fig:bulk}
\end{figure*}

\begin{table}
\centering
\small
  \caption{Experimental\cite{Jonsson1976,maloney2014} and computed (at HF-3c level of theory) cell parameters of ethanol and ethylamine bulk structures. \textbf{a}, \textbf{b} and \textbf{c} are in \AA, $\alpha$, $\beta$ and $\gamma$ are in degrees, and the volumes are in \AA$^3$.}
  \label{tab:cell_param}
  \begin{tabular*}{0.99\columnwidth}{@{\extracolsep{\fill}}lcccc}
  & \multicolumn{2}{c}{Ethanol} & \multicolumn{2}{c}{Ethylamine} \\
   & Exp. & Comput. &  Exp. & Comput. \\
   \hline
    \textbf{a}  &5.377 & 5.179  & 8.263 &    7.856  \\	
     \textbf{b}  &6.882 & 6.773   & 7.310 &     6.949  \\	
     \textbf{c}  & 8.255 & 8.146  & 5.532 &    5.315  \\	
     $\alpha$ & 90.0 & 90.0   & 90.0 &    89.9 \\ 
     $\beta$ & 102.2 & 104.8  & 99.1 &    100.0 \\ 
     $\gamma$ & 90.0  & 90.0  & 90.0 &    89.8 \\ 
     Volume & 298.6  & 276.2  & 329.9 &   285.8 \\
  \hline
  \end{tabular*}
\end{table}
The bulk and surface ice models for EtOH and EtNH$_2$ were simulated taking as a reference their experimental structures. The crystal structure of EtNH$_2$ (P2$_1$/C space group) was determined by X-ray diffraction at 150 K through \textit{in situ} crystallization from the liquid,\cite{maloney2014} while single crystals of ethanol (P$_C$ space group) were grown at 156 K and analyzed through X-ray diffraction at 87 K.\cite{Jonsson1976}

In the asymmetric unit, both EtOH and EtNH$_2$ present two crystallographically independent molecules, while the unit cells contain four molecules (see Figures \ref{fig:bulk}a and \ref{fig:bulk}d). In both cases, two types of intermolecular interactions can be distinguished: H-bonding between the NH$_2$ (for EtNH$_2$) and OH (for EtOH) groups, and CH$_3$···CH$_3$ dispersion. The conformation around the \ch{C\bond{single}C} bond is staggered in both molecules. In EtOH, the OH group is differently oriented so that both the \textit{trans} and the \textit{gauche} conformers are present in the unit cell.
The optimization at HF-3c level of theory results in a reduction of the cell parameters, and consequently of the cell volume by 7\% for EtOH and by 13\% for EtNH$_2$ (see Table \ref{tab:cell_param}).

The (010), (100), (001), (110), (101), (011), and (111) surfaces were generated from the EtOH and EtNH$_2$ bulk structures, the thicknesses of which satisfy the surface energy convergence condition.\cite{corno2014} Special care was paid to the resulting dipole moment across the non-periodic direction of each slab, discarding those with |$\mu$| $>$ 5 D. The stability order of the surfaces was determined primarily on their E$_S$, and in second instance on their E$_C$ and E$_X$ (shown in Table \ref{tab:surfaces}). For both EtOH and EtNH$_2$, the slabs present very small E$_S$, in agreement with their weakly bounded molecular crystal nature. 
The extraction energy was computed for each molecule laying in the external layer of each surface. Thus, two to four values were computed for each slab. The (100) slabs represent the only exception for both systems, as the molecules defining the top layer are related by symmetry (a glide reflection in EtOH, and a screw rotation in EtNH$_2$) hence providing only one value.  

Six EtOH surfaces (see Table \ref{tab:surfaces}) were found complying with the dipole moment constraint. The ladder of stability is (from more to less stable): (010)$^{A-layer}$ > (100) > (101) > (110) > (011) > (010)$^{B-layer}$ (the definition of A-layer and B-layer is explained below). The cohesive energies obtained for these slabs ranges from -49.4 kJ mol$^{-1}$ (010)$^{A-layer}$ to -40.3 kJ mol$^{-1}$ (011), while the extraction energy spans the range from 59 to 115 kJ mol$^{-1}$.

For EtNH$_2$, eight surfaces were characterized (see Table \ref{tab:surfaces}), and their stability order is: (100)$^{A-layer}$ > (001) > (010) > (111) > (101) > (110) > (011) > (100)$^{B-layer}$. All of them have an almost null dipole, with the exception of the (110) slab ($\mu$ = 1.8 D), which is responsible for the large E$_S$.  
The E$_C$ values range between -40.4 kJ mol$^{-1}$ (100)$^{A-layer}$ and -30.0 kJ mol$^{-1}$ (011), and E$_X$ covers the range between 35 and 85 kJ mol$^{-1}$, which makes ethylamine slightly less bound than ethanol.

Two types of cuts (A-layer and B-layer) were modeled for the (010) EtOH and the (100) EtNH$_2$ surfaces, the former one being the most stable slab cut. 

In the EtOH-(010) and EtNH$_2$-(100) surfaces, a couple of layers hold together by a network of H-bonds is alternated to another couple governed by dispersive forces. Depending on the type of interlayer interaction preserved by the cut, the resulting surface can be characterized by the polar moieties being completely segregated inwards to the slab (A-layer, see Figures \ref{fig:bulk}b and \ref{fig:bulk}e) or totally exposed to the exterior (B-layer, see Figures \ref{fig:bulk}c and \ref{fig:bulk}f). The presence of polar groups exposed to the surface causes the B-layer cut to be less stable than the A-layer, resulting in EtOH-(010)$^{B-layer}$ and EtNH$_2$-(100)$^{B-layer}$ being characterized by the largest E$_S$. The natural consequence is that B-layer type surfaces will yield larger BE(0)s for the adsorption of \ch{CH3CH2OH} and \ch{CH3CH2NH2}. 


\begin{table*}[htb]
\centering
\small
  \caption{Dipole moment across the slab ($\mu$, in Debye), surface energy (E$_S$, in J m$^{-2}$), cohesive energy (E$_C$, in kJ mol$^{-1}$), and extraction energy (E$_X$, in kJ mol$^{-1}$) of EtOH and EtNH$_2$ surfaces at B3LYP-D3(BJ)/A-VTZ* level of theory. E$_C$, E$_S$, and E$_X$ are corrected for the ZPE.}
  \label{tab:surfaces}
  \begin{tabular*}{0.75\textwidth}{@{\extracolsep{\fill}}lllll}
\multicolumn{5}{c}{EtOH}  \\
  Surface  & $\mu$ &  E$_S$ & E$_C$  & E$_X$  \\
    \hline
(010)$^{A-layer}$ & +7.2 $\times$ 10$^{-4}$  &		0.0761 & -49.4 &		104.1 -- 108.2 \\  
(100) & $-$3.3 & 0.0994 &  -47.9 &  106.0 \\ 
(101) & $-$1.7 & 0.1053 &  -45.7 &  62.0 -- 113.7 \\  
(110) & $-$2.9  & 0.1594 &  -44.0 &  59.0 -- 116.6 \\  
(011) & +3.5   & 0.1616 &  -40.3 &  60.0 -- 95.2 \\  
(010)$^{B-layer}$ & $-$1.8 $\times$ 10$^{-1}$ &    0.1857 & -42.0 &   58.9 -- 60.0 \\
    \hline
    \\
\multicolumn{5}{c}{EtNH$_2$} \\
  Surface  & $\mu$  &  E$_S$  & E$_C$  & E$_X$  \\
    \hline
(100)$^{A-layer}$ &	+1.3 $\times$ 10$^{-4}$    & 	0.0673 & -40.4 &		71.5 \\ 
(001) &	+2.4 $\times$ 10$^{-3}$    &		0.0797 &  -39.9  &		64.8 -- 69.3 \\  
(010) &	+2.5 $\times$ 10$^{-4}$    &		0.0790 &  -38.9  &		65.0 -- 69.4 \\  
(111) & $-$1.2 $\times$ 10$^{-3}$   &     0.0962 &     -35.7 &      63.3 -- 80.4 \\  
(101) &	+1.2 $\times$ 10$^{-3}$     &		0.1034 &   -33.8 &		36.0 -- 63.5 \\  
(110) &  +1.8            & 0.1295 &    	-36.7 &       68.1 -- 84.3 \\  
(011) & +8.8 $\times$ 10$^{-4}$	  & 		0.1319 & -30.0 &		37.5 -- 60.4 \\  
(100)$^{B-layer}$ & $-$1.5 $\times$ 10$^{-3}$ &    0.1857 & -34.1 &   45.6 \\
\hline
\end{tabular*}
\end{table*}

\subsubsection{Binding energies on pure \ch{EtOH} and \ch{EtNH2} surfaces}\label{par:be_icom}

We computed three BE(0)s for each species, two on the  \ch{EtNH2}-(100)$^{A-layer}$ and \ch{EtOH}-(010)$^{A-layer}$ surfaces, and one on the \ch{EtNH2}-(100)$^{B-layer}$ and \ch{EtOH}-(010)$^{B-layer}$ surfaces. The structures are available in the ESI$^\dag$.

On the EtOH-(010)$^{A-layer}$ surface, we identified two adsorption complexes characterized by a different orientation of the adsorbate with respect to the surface. In the first case, the driving force of the interaction is represented by the dispersive interactions between the aliphatic chains, resulting in the ethanol molecule laying parallel to the surface, with BE(0) = 20.9 kJ mol$^{-1}$. In the second case, the molecule is placed perpendicularly to the surface to establish a H-bond interaction with one of the partially exposed hydroxyl groups. Dispersive forces also play an important role in the BE(0), which is as high as 30.9 kJ mol$^{-1}$. 

On the \ch{EtNH2}-(100)$^{A-layer}$ surface, we found two very similar adsorption complexes with a BE(0) of 21.4 kJ mol$^{-1}$ and 21.6 kJ mol$^{-1}$, due to the symmetry of the surface. 
They are entirely dominated by dispersive forces, the adsorbate molecule laying parallel to the surface in the two cases. The small difference between the two complexes is due to a slight change in the orientation of the ethylamine with respect to the EtNH$_2$ surface. 

The adsorption complex obtained on the \ch{EtNH2}-(100)$^{B-layer}$ and \ch{EtOH}-(010)$^{B-layer}$ slabs are characterized by the \ch{-OH} and \ch{-NH2} groups of the adsorbate accepting an H-bond from the surface, resulting in BE(0) = 41.8 kJ mol$^{-1}$ for \ch{CH3CH2OH} and BE(0) = 43.0 kJ mol$^{-1}$ for \ch{CH3CH2NH2}. 

\subsubsection{Binding energies on the crystalline and amorphous water ice models}\label{par:be_ice}

The theoretical results on the single adsorption (S-ADS) of \ch{CH3CH2OH} and \ch{CH3CH2NH2} on the crystalline ice (CI) and amorphous solid water (ASW) surfaces are presented here.

\begin{figure*}[htb]
\centering
  \includegraphics[width=0.9\textwidth]{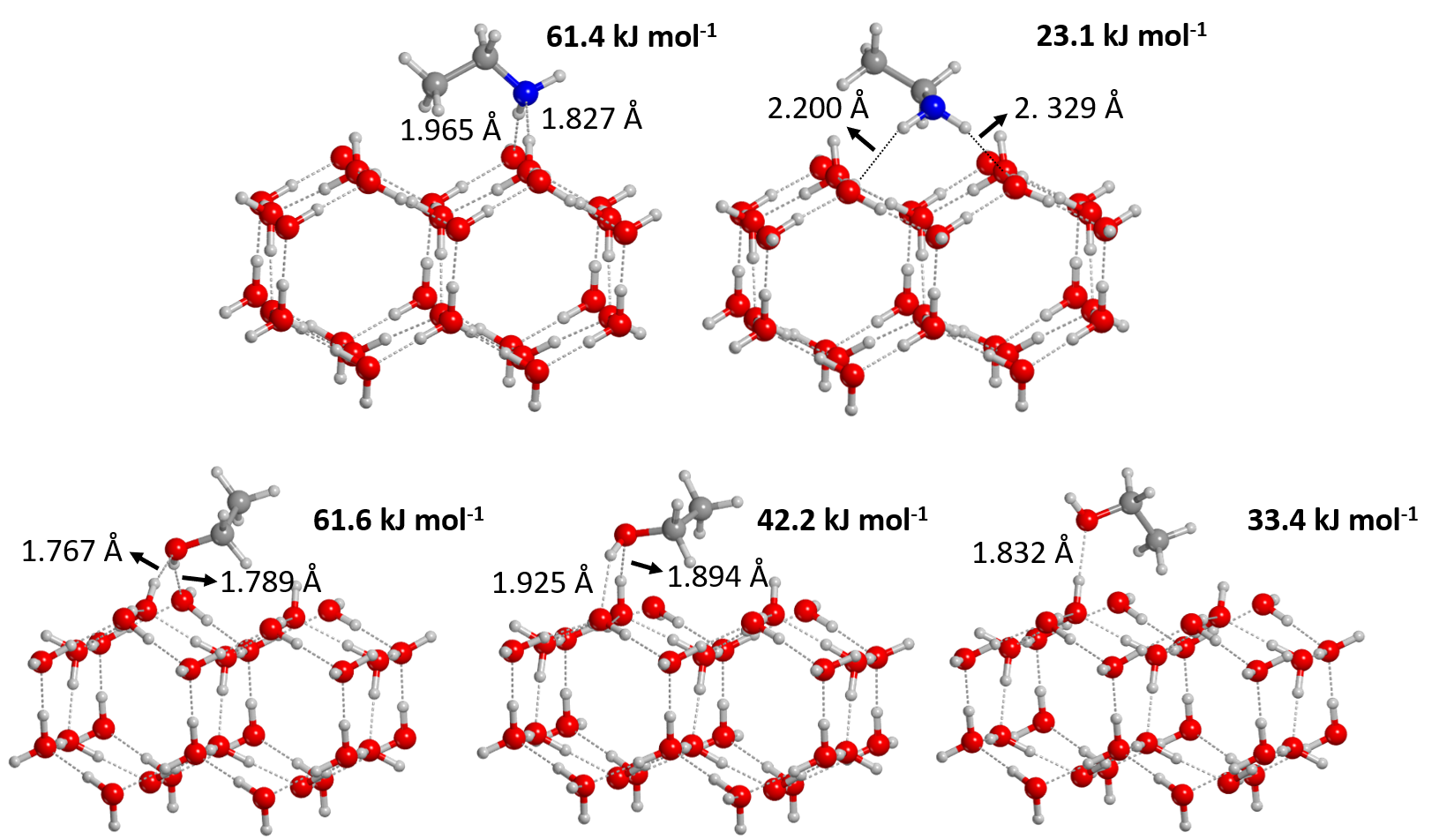} 
  \caption{Optimized geometries of the adsorption complexes of ethylamine (top) and ethanol (bottom) on the crystalline water ice surface model. The corresponding computed BE(0)s are also given.}
  \label{fig:crystalline}
\end{figure*}

We identified three adsorption complexes of \ch{CH3CH2OH} onto the CI surface, characterized by a different orientation of the adsorbate with respect to the surface (see bottom panel of Figure \ref{fig:crystalline}). Each complex is characterized by the presence of at least one H-bond. In the weakest bound complex (BE(0) = 33.4 kJ mol$^{-1}$), \ch{CH3CH2OH} only acts as H-bond acceptor, in favour of the interaction of the aliphatic chain with the ice. In the other two complexes, both the O and H atoms of \ch{CH3CH2OH} are involved in H-bonds, but the orientation of the adsorbate, in the most stable case (BE(0) = 61.6 kJ mol$^{-1}$), precisely interacts with uncoordinated surface-exposed H$_2$O sites, while, in the other (BE(0) = 42.2 kJ mol$^{-1}$), a rearrangement of a surface dangling OH groups is required to establish two (longer) H-bonds. 

For \ch{CH3CH2OH} on the ASW surface, nine binding sites were sampled. The majority of the BE(0) values ranges from 37 to 48 kJ mol$^{-1}$ (see Table \ref{tab:water}), with two exceptions. In dO8 site, the ethanol OH group only acts as a (weak) H-bond donor towards the dO atom and hence BE(0) = 26.0 kJ mol$^{-1}$. In contrast, in dH7 site, ethanol is located in the cavity of the ASW and establishes three H-bonds, two involving the O atom and one the H atom of the OH moiety, and hence BE(0) = 59.1 kJ mol$^{-1}$.

The adsorption of \ch{CH3CH2NH2} on the CI surface model resulted in two adsorption complexes, each characterized by the presence of two H-bonds (see top panel of Figure \ref{fig:crystalline}). In the most stable complex, \ch{CH3CH2NH2} is accepting a H-bond through the N atom and donating another H-bond through a H atom of the amino group, resulting in BE(0) = 61.4 kJ mol$^{-1}$, where the contribution of weak forces is reduced to 40\%. In the second (less stable) complex, the two H atoms of the amino group act as H-bond donors, thus resulting in a weaker complex with BE(0) = 23.1 kJ mol$^{-1}$), where the dispersion interactions of the aliphatic chains are responsible for the larger contribution (almost 90\%) to the final BE(0). 

For \ch{CH3CH2NH2} on the ASW surface, similarly to what was observed for \ch{CH3CH2OH}, the nine BE(0)s range from 45 to 55 kJ mol$^{-1}$, with the exception of those computed for sites dO8 and dH7, with BE(0) = 19.2 kJ mol$^{-1}$ and 71.5 kJ mol$^{-1}$, respectively. The reason is also a weak H-bond donation for the first case and a very strong interaction dominated by three H-bonds involving each atom of the \ch{NH2} group for the second case.

\begin{table}[htb]
\centering
\small
  \caption{Computed binding energies (BE(0), in kJ mol$^{-1}$) of ethanol and ethylamine on the ASW ice surface model. See Figure \ref{fig:ice_models} for the definition of the binding sites.}
  \label{tab:water}
  \begin{tabular*}{0.75\columnwidth}{@{\extracolsep{\fill}}lcc}
Binding Site& \ch{CH3CH2OH} & \ch{CH3CH2NH2} \\
\hline
dH1 & 37.0 & 50.1 \\
dH2 &  44.8 & 47.8 \\
dH3 &  40.3 & 54.0 \\
dH4 &  42.0 & 54.2 \\
dH5 & 40.5 & 44.8 \\
dH6 &  47.8 & 50.2 \\
dH7 & 59.1 & 71.5 \\
dO8 &  26.0 & 19.2 \\
dO9 &  37.1 & 46.2 \\
\hline
  \end{tabular*}
\end{table}

\subsubsection{Half and full monolayers on crystalline water ice}\label{par:be_monolayer}

The simulation of higher surface coverages consisted in the adsorption of two (H-ML) or four (F-ML) molecules per cell. The adsorbates were manually placed with the \ch{NH2} and \ch{OH} moieties pointing towards the surface to optimize their interaction with the ice.

For both species, the H-ML BE(0) per molecule are slightly larger than that obtained in the S-ADS regime: 66.0 kJ mol$^{-1}$ (versus 61.6 kJ mol$^{-1}$) for \ch{CH3CH2OH}, and 64.5 kJ mol$^{-1}$ (versus 61.4 kJ mol$^{-1}$) for \ch{CH3CH2NH2}. Instead, the F-ML BE(0) per molecule are smaller, 50.0 kJ mol$^{-1}$ for  \ch{CH3CH2OH} and 49.5 kJ mol$^{-1}$ for \ch{CH3CH2NH2}. This is due to the fact that, in the H-ML cases, two molecules can still easily accommodate in the cell and find the most favourable geometry to maximize their interactions with the surface. In contrast, in the F-ML cases, the four molecules struggle to fit in the cell and lay with the aliphatic chain pointing upwards, minimizing the interaction with the ice and being surrounded only by the other adsorbates. The lateral interaction between the adsorbates, one of the terms favourably contributing to the BE(0), is mostly based on weak London forces, increasing the weight of dispersive interactions in the final BE(0): from the 43\% and 44\% in H-ML conditions, to the 68\% and 79\% in the F-ML ones, for \ch{CH3CH2OH} and \ch{CH3CH2NH2}, respectively.

Despite the compromises made by our models and the simplified procedures, the present results are guidelines for understanding the process of approaching a monolayer coverage in a real system. That is, the first adsorbate molecules occupy the strongest surface adsorption sites, and then, by increasing the number of adsorbates, surface sites from the strongest to the weakest ones become occupied, until the formation of a monolayer. This assumption holds true in case of molecules that wet the surface, like ethanol and ethylamine.

\section{Laboratory Experiments}\label{sec:experiments}

The desorption energies (E$_{\rm des}$) of \ch{CH3CH2OH} and \ch{CH3CH2NH2} have been estimated by means of experimental procedures, dividing the work in two parts: determination of the monolayer (ML) of ethanol and ethylamine, and temperature program desorption (TPD) measurements to determine the E$_{\rm des}$ of the adsorbates on crystalline and amorphous water ice films.  

\subsection{Methodology}

\subsubsection{Experimental setup}

All of the hereby experiments were conducted with the VENUS apparatus located at LERMA-CY laboratory in Cergy-Pontoise, France, and described extensively in \citet{Congiu2020}. VENUS consists of an ultra-high vacuum (UHV) stainless steel main chamber with a base pressure of 2 $\times$ 10$^{-10}$ mbar, containing a 9 mm in diameter gold-coated copper sample holder. The latter is  attached to the cold head of a closed-cycle He cryostat, that allows to vary its temperature between 6 K and 350 K via a computer-controlled resistive heater. 

Ethanol is liquid at room temperature, while ethylamine needs to be diluted in water to be stable (with about 66--72\% of \ch{CH3CH2NH2} in \ch{H2O} \footnote{Ethylamine; SDS No. 295442 [Print]; Merck Life Science S.A.S.: Saint-Quentin-Fallavier, France, Feb 16, 2023. \url{https://www.sigmaaldrich.com/FR/fr/product/aldrich/471208} (accessed Nov 9, 2023)}). The two species were first converted into gas in the central molecular beamline of VENUS. The beam has an aperture of approximately 3 mm and a residual pressure of 10$^{-7}$ mbar. By means of a differential pumping through two intermediate stainless steel chambers, the gases were then sent towards the cold surface. The species were deposited on the surface maintained at a constant temperature of 70~K. During this process, the base pressure of the main chamber remains low (up to 5 $\times$ 10$^{-10}$ mbar), ensuring a focalization of the molecules onto the sample and negligible gas-gas interactions above the cold head. During this phase, it is possible to monitor the evolution of the ice \textit{in situ} via reflection absorption infrared spectroscopy (RAIRS) performed with a Vertex-70 Fourier Transform infrared (FT-IR) spectrometer over a spectral range from 800 to 4500 cm$^{-1}$ (from 12.5 to 2.22 nm). The production of a mean spectrum is non-destructive and takes place every two minutes, corresponding to 256 scans of the ice. 

Once the deposition phase is completed, the desorption of the species from the surface is monitored via temperature programmed desorption (TPD). After positioning the Hiden 51/3F Quadrupole Mass Spectrometer (QMS) 5 mm in front of the cold head, the surface temperature is gradually increased with a rate of $\beta$ = 0.15 K s$^{-1}$ = 9 K min$^{-1}$ up to 200 K. Throughout the heating process, the different chemical species progressively sublimate from the surface and desorb at a temperature that depends on their nature and on the binding energy with each site offered by the surface. This results in a TPD spectrum where the abundance of  each atomic mass unit (amu) is monitored against its thermal desorption temperature. Indeed, the electrons produced by the hot filament of the QMS will likely induce the fragmentation of the molecules hitting its head into several ions. The statistic repartition of the ions, called cracking pattern, is specific for each species and depends on the ionization energy (set at 30 eV in this work). Each desorption curve can be described with the Polanyi--Wigner equation following an Arrhenius law.
\begin{equation} 
    \centering
    \label{<Polanyi_Eq>}
    r(N, E_{\rm des}, T) = -\frac{\mathrm{d}N}{\mathrm{d}t} = \frac{A}{\beta} N^{n} e^{\frac{-E_{\rm des}}{RT}}
\end{equation}
where the desorption rate \textit{r} (molecules cm$^{-2}$ s$^{-1}$) depends on the number density of molecules adsorbed on the surface \textit{N} (molecules cm$^{-2}$), the desorption energy of a molecule on the surface E$_{\rm des}$ (J mol$^{-1}$), and the temperature of the surface \textit{T} (K). In equation \ref{<Polanyi_Eq>}, \textit{A} is the pre-exponential factor (s$^{-1}$), $\beta$ is the heating rate, R is the ideal gas constant (J mol$^{-1}$ K$^{-1}$), and \textit{n} is the order of the desorption process. 
The zeroth order represents a desorption kinetics that is independent of the amount of available adsorbate, such as the case of thick films consisting of multilayers. 
The first-order kinetics corresponds to the desorption of an adsorbate whose coverage is lower or equal to one monolayer, and implies that the desorption rate is proportional to the number of molecules present on the surface. The fitting of the TPD allows us to determine \textit{N} and \textit{E${_{\rm des}}$}, provided that the pre-exponential factor \textit{A} is fairly constrained.

\begin{table*}[htb]
\centering
\small
\caption{List of the experiments performed with the VENUS setup.}
\label{tab:list_exp}
\begin{tabular}{llll}
    Deposited molecule & Quantity deposited (ML) & Substrate type & Substrate thickness (ML) \\
    \hline
    Ethanol (\ch{C2H5OH}) & 0.25 ; 0.5 ; 0.6 ; 0.7 ; 0.75 ; 1.5 & Gold & - \\
      & 0.15 ; 0.25 ; 0.7 ; 2.1 & Amorphous \ch{H2O} & 10 ; 6.5 ; 10 ; 10 \\
      & 0.15 & Crystalline \ch{H2O} & 10 \\
    \hline
    Ethylamine (\ch{C2H5NH2}) & 0.2 ; 0.33 ; 0.4 ; 0.7 ; 0.83 ; 1.25 & Gold & - \\
      & 0.08 ; 0.13 ; 0.4 ; 1.25 & Amorphous \ch{H2O} & 10 ; 10 ; 10 ; 10 \\
      & 0.13 & Crystalline \ch{H2O} & 10 \\
    \hline
\end{tabular}
\end{table*}

\subsubsection{Growth of the molecular films}
\label{sss_calib_molecular_films}

In this work, ethanol and ethylamine have been sent towards the surface with the central molecular beam of VENUS. They have been deposited either on bare gold or on a film of water ice (see section \ref{sss_water_growth} for more details). In all the cases, the surface was held at 70~K. The VENUS setup has been optimized for the study of very thin layers (from sub-monolayer up to one monolayer) of adsorbates, although systems of thicker ices, whether pure or mixed, can be investigated as well (usually under 100 monolayers). In order to determine the amount of molecules present on the cold head, two methods can be used and mutually confirm each other.

The first method can be performed \textit{in situ} via infrared measurements, during the deposition. The column density $N$ (molecules~cm$^{-2}$ ) of deposited or newly formed species can be calculated via a modified Lambert-Beer equation,
\begin{equation} 
N = 2 \, \frac{\int A(\lambda) \mathrm{d}\lambda}{f},
    \label{<Lambert_Eq>}
\end{equation}
where the constant 2 is determined specifically for the VENUS setup,\citep{Congiu2020} $\int A(\lambda) \, \mathrm{d}\lambda$ is the integrated area of the infrared absorption feature (cm$^{-1}$), and $f$ is the corresponding band strength. In the case of ethanol, the monolayer is calculated to be reached after approximately 21 minutes, using a band strength of $f = 1.41 10^{-17}$ cm $\times$ molecules$^{-1}$ at 1055 cm$^{-1}$.\cite{hudson_2017} For ethylamine, the only band strength values available have been calculated very recently.\citep{Hudson2022} As there are no other available values, most experimental works involving ethylamine (i.e., \citet{Danger2011}) adopted the band strength of methylamine, which was first communicated in 2005 without any known source.\citep{Holtom2005} With the corresponding band strength of amorphous ethylamine $f = 3.69 10^{-18}$ cm $\times$ molecules$^{-1}$ at 1395 cm$^{-1}$,\citep{Hudson2022} one monolayer of \ch{CH3CH2NH2} appears to be reached after 24 minutes of deposition. 

The other way is via the TPD measurements as described in \citet{Noble2011}. A monolayer (ML) is defined by the theoretical filling of all available sites on the gold surface, corresponding to approximately 10$^{15}$ molecules cm$^{-2}$. 
During the deposition phase, as the dose is gradually increased, the available sites are progressively filled beginning from high-depth sites towards low-energy ones, until the deposited amount reaches 1~ML.
In the case of ethanol, the values obtained with TPD and infrared measurements match well, giving a ML completion after $\sim$20 minutes. This is easily determinable with the desorption profile of \ch{CH3CH2OH}. Ethylamine, however, has a more complex desorption profile, making the monolayer determination more ambiguous. The areas under the main fragment curves of ethylamine and ethanol are not directly comparable, due to their different ionization cross-section at 30 eV. At this ionization energy, there is no record of an ionization cross-section measurement for ethylamine. The value of $\sigma_{\ch{CH3CH2NH2}}^{\rm tot}$ = 9~\AA$^{2}$ has been used in studies of 67P/C-G \citep{goesmann2015,Altwegg2016} but is only valid at 70 eV. In this work, we chose to use the ionization cross-section of dimethylamine ($\sigma_{\ch{CH3NHCH3}}^{\rm tot}$ = 6.68~\AA$^{2}$ at 30 eV), which also has a 9~\AA$^{2}$ value at 70 eV.\citep{Singh2018} Considering this, it is possible to demonstrate once again the time of ML completion for ethanol (20 minutes) and ethylamine (24 minutes). Table \ref{tab:list_exp} reports the list of experiments performed.

\subsubsection{Growth of water ice films} \label{sss_water_growth}

Water can be grown on the sample either via a molecular beam identical to the one used for ethanol and ethylamine, or via a water vapor delivery manifold controlled by a needle valve. In this work, the latter technique was used. It allows to leak water vapor directly into the main chamber, sufficiently far from the cold head for a monolayer to be grown in approximately 5 minutes. This method is also called ``background deposition'' and enables the possibility to use water as a matrix, or as a thick ice substrate. In the experiments of this work, we deposited 10 ML of water on the sample, with the exception of one experiment, in which a more accurate recalculation of the ice thickness yielded 6.5 ML (see last column of Table \ref{tab:list_exp}).

Depending on the temperature and modality of water deposition, different morphologies can be obtained. In our experimental conditions, non-porous amorphous solid water (ASW) - mimicking a compact ice bulk - is easily formed by leaving the cold finger at a steady temperature of 110 K during the deposition. Once a thickness of 10 ML is reached, the sample is cooled down to 70 K in order to deposit one of the molecules of interest. The procedure to form crystalline ice (CI) requires more steps, as a thermal treatment has to be performed onto the 10 ML formerly deposited. The ice is first warmed to 140 K at a rate of 9 K min$^{-1}$, then to 142.5 K at a rate of 6 K min$^{-1}$. The process of crystallization can be monitored via RAIRS (stopped once the entire ice is crystalline, proved by a major change in the band shape), as well as via QMS to quantify the desorbing \ch{H2O} until this value remains constant.

The quantity of \ch{H2O} on the surface can directly be estimated during the deposition by monitoring the partial pressure of water inside the main chamber. Otherwise, it can be verified using the same methods as for ethanol and ethylamine.

\subsection{Results}

The experiments performed in this work can be divided into two categories: i) sub-monolayer depositions of the adsorbates on ASW or CI films; and ii) multilayer depositions of the adsorbates on an ASW film. For each adsorbate, \ch{CH3CH2OH} and \ch{CH3CH2NH2}, three sub-monolayer depositions were carried out on ASW and one on CI. Only one multilayer deposition of each species was performed on amorphous \ch{H2O}. See Table \ref{tab:list_exp} for the experimental details.

With this experimental study, our aim is to investigate the modification of the desorption profiles of \ch{CH3CH2OH} and \ch{CH3CH2NH2} depending on the surface type on which they are deposited (gold, ASW and CI), and on the surface coverage (sub-monolayer, monolayer and multilayer). In order to compare these results with the theoretical ones, the desorption energy of the adsorbates was also derived from the TPD curves using an in-house developed software.

In the figures of this section, for the sake of clarity, we only display the most abundant molecular ion for each species: \textit{m/z} = 18 for water ([\ch{H2O}]$^{+}$), \textit{m/z} = 31 for ethanol ([\ch{CH2OH}]$^{+}$), and \textit{m/z} = 30 for ethylamine ([\ch{CH2NH2}]$^{+}$). Moreover, as no clear conclusion can be drawn from the infrared spectra produced during the experiments, except for the monolayer calibration (see subsection \ref{sss_calib_molecular_films}), we chose not to display them in this paper.

\subsubsection{Sub-monolayer depositions}
\label{sss_submono_dep}

\begin{figure}[ht]
\centering
  \includegraphics[width=\columnwidth]{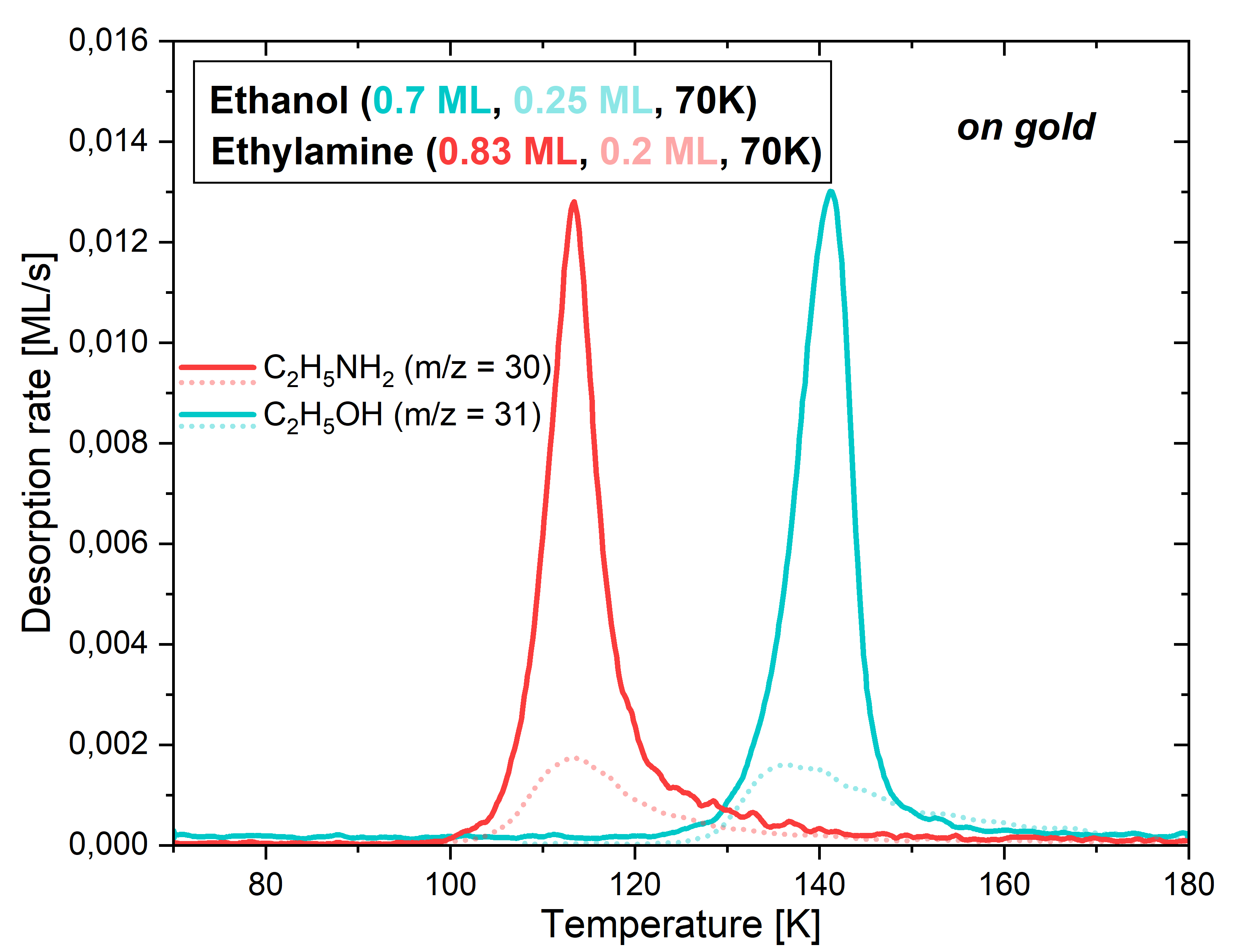} \\
  \caption{TPD curves for the deposition of 0.7 ML (solid blue line) and 0.25 ML (dotted clear blue line) of ethanol, and 0.83 ML (solid red line) and 0.2 ML (dotted clear red line) of ethylamine, on the gold substrate. The most abundant molecular ion is displayed for each molecule.}
  \label{fig:gold_etoh_etnh2}
\end{figure}

\paragraph{On a gold substrate}
$\,$
\newline
Firstly, the behaviour of \ch{CH3CH2OH} and \ch{CH3CH2NH2} on the bare gold surface was studied to assess the adsorbate/surface interactions, which will be compared with those established when the surface consists of water ice (described in the following subsections). The experiments were initially performed to calibrate the flux to use for each of the two adsorbates. Figure \ref{fig:gold_etoh_etnh2} represents the TPD curves for the \textit{m/z} = 31 ethanol fragment and the \textit{m/z} = 30 ethylamine fragment, which desorb at different deposition times. To avoid any ambiguity, and since the deposition times account for the quantities deposited, they have been converted to monolayers (MLs).

While \ch{CH3CH2OH} exhibits a noticeable site-filling behaviour, starting with the strongest binding sites, this is not the case for \ch{CH3CH2NH2}.
The start, end, and maximum of the desorption occurs at 122, 170 and 141 K for 0.7 ML of ethanol, and at 97, 160 and 113 K for 0.83 ML of ethylamine. In the case of the lowest coverages, represented by dotted and more transparent lines, the start of the desorptions is slightly shifted towards higher temperatures. This is explained by the propensity of the adsorbates to attach the most favourable binding sites, therefore needing more thermal energy to sublimate into the gas phase. As soon as more molecules are present on the surface, they occupy less favourable binding sites, until all of them are used. This moment defines the monolayer, as additional molecules would directly stick to the first layer, and not to the bare gold. It is worth noting that the calibration curve pattern matches the time estimated for the monolayer completion in the previous section. 

In these conditions, it is possible to extract the desorption energy of each of the adsorbates on the gold substrate, following the method outlined in subsection \ref{sss_BE_derivation}.

\begin{figure*}[ht]
\centering
  \includegraphics[width=0.9\textwidth]{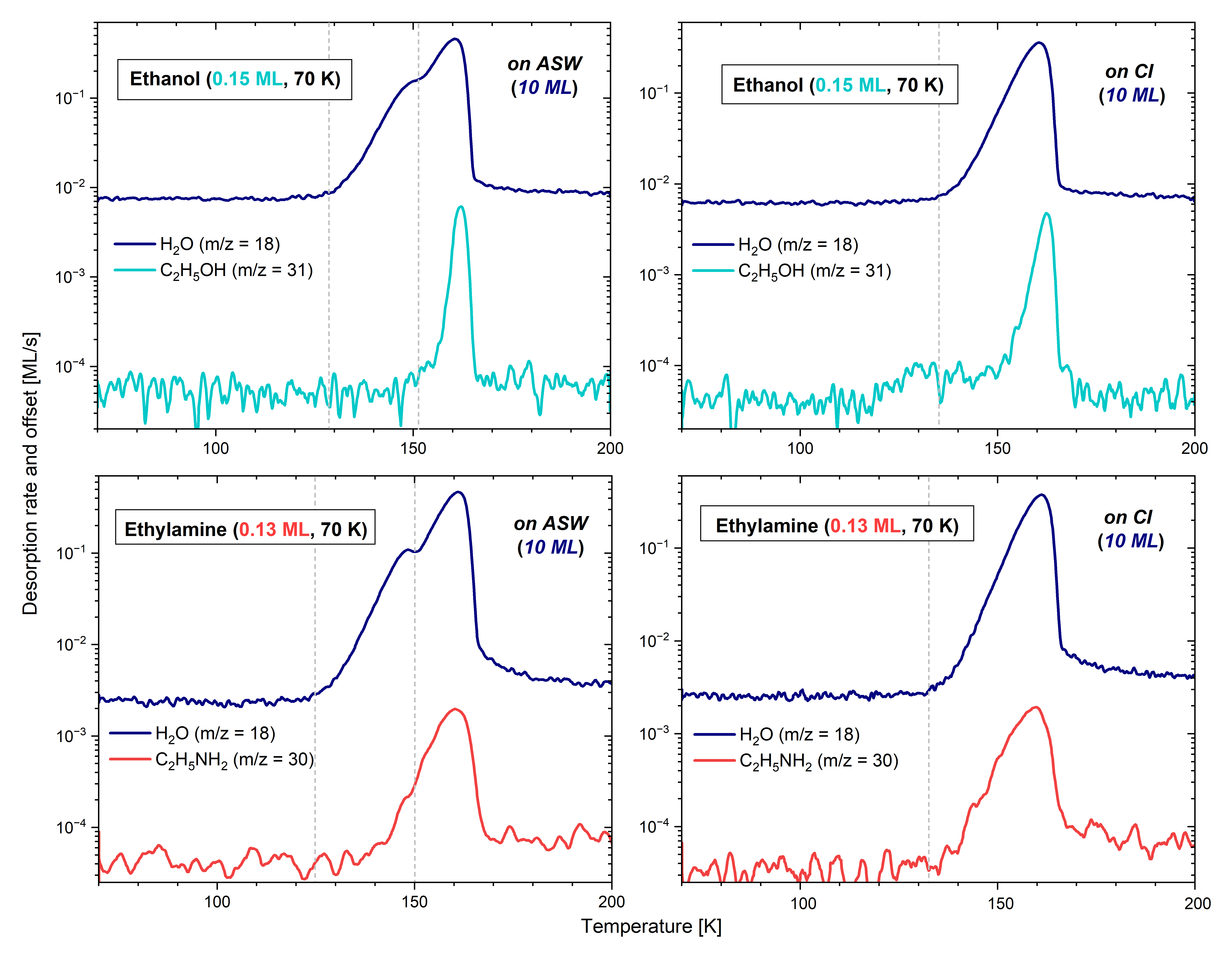} \\
  \caption{TPD curves for the deposition of 0.15 ML of ethanol on 10 ML of ASW (top left) and of CI (top right), and 0.13 ML of ethylamine on 10 ML of ASW (bottom left) and of CI (bottom right). The most abundant molecular ion is displayed for each molecule. 
  A logarithmic scale is used for a better data readability, and grey vertical lines indicate the start of the water desorption and its crystallization.}
  \label{fig:asw_ci_etoh_etnh2_0.15}
\end{figure*}

\begin{figure*}[ht]
\centering
  \includegraphics[width=0.9\textwidth]{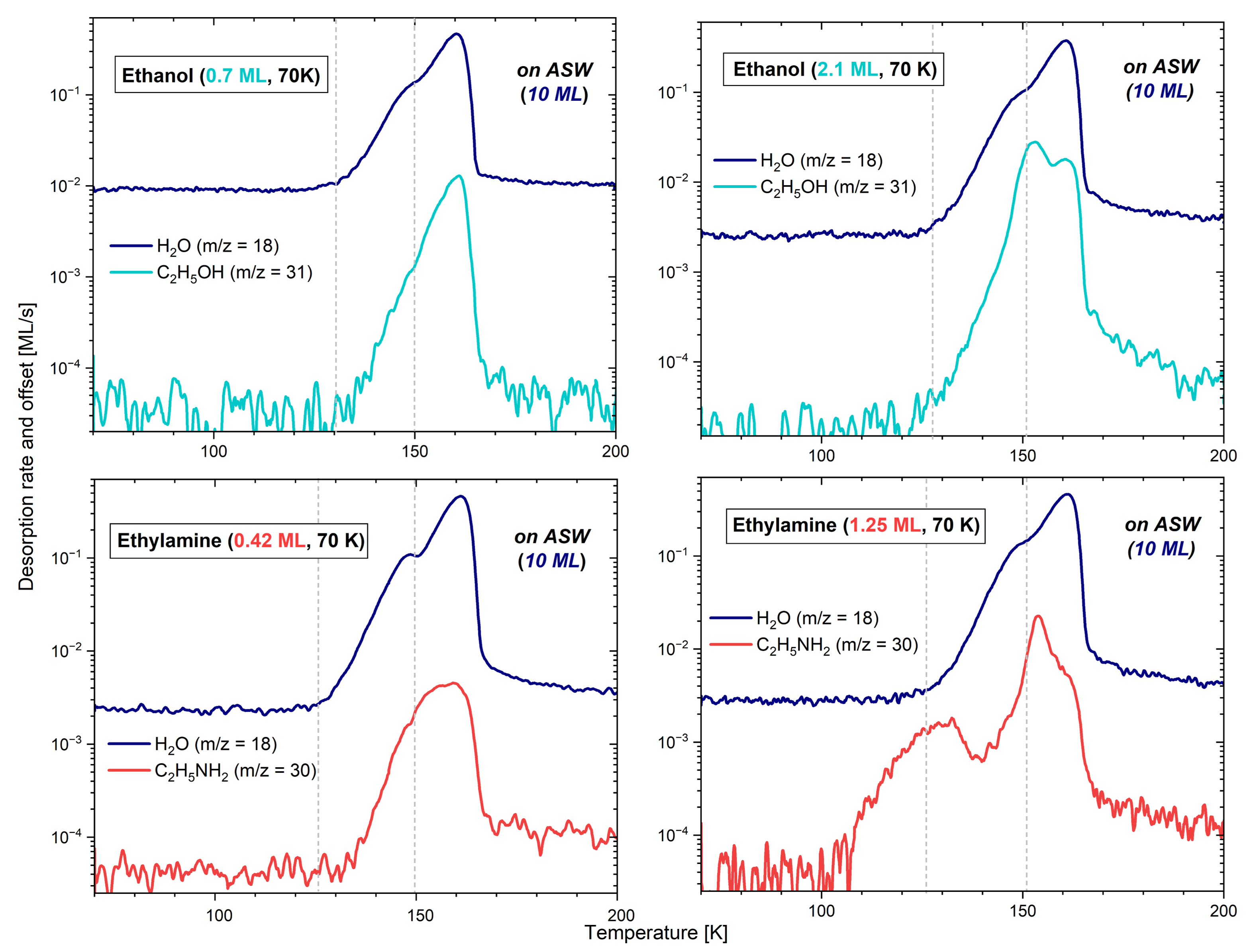} \\
  \caption{TPD curves for the deposition of 0.7 ML of ethanol (top left), 0.4 ML of ethylamine (bottom left), 2.1 ML of ethanol (top right) and 1.25 ML of ethylamine (bottom right) on 10 ML of ASW. The most abundant molecular ion is displayed for each molecule. A logarithmic scale is used for a better data readability, and grey vertical lines indicate the start of the water desorption and its crystallization. 
 }
  \label{fig:asw_etoh_etnh2_submono_multi}
\end{figure*}

\paragraph{On amorphous solid water}
$\,$
\newline
In the case of the sub-monolayer depositions of \ch{CH3CH2OH}, namely 0.15, 0.25 and 0.7 ML, on a slab of 10 ML of ASW, the desorption curves and behaviours are all fairly comparable, as seen on the left panel of Figures \ref{fig:asw_ci_etoh_etnh2_0.15} and \ref{fig:asw_etoh_etnh2_submono_multi}. 

The TPD profiles differ in their integrated area, the latter being proportional to the quantity of molecules deposited on the bulk of \ch{H2O}. In the case of the 0.15 and 0.25 ML \ch{CH3CH2OH} depositions, a common pattern is visible, with the adsorbate starting to desorb at 147 K and 144 K, respectively, along with the crystallization of water. On the other hand, with the 0.7 ML \ch{CH3CH2OH} deposition, the adsorbate starts to desorb together with amorphous water, already at $\sim$130 K. We note here that the quantity of water under the 0.25 ML deposition of ethanol was lower due to the use of a lower flux. The additional noise, visible on \textit{m/z} = 18 and \textit{m/z} = 31, is due to a poor following of the heating ramp after a change of the thermometer. This resulted in small recurrent surges of desorption, and in the partial crystallization of water before the expected temperature. However, this behaviour does not change the essence of the results. In all cases, the rest of the ethanol desorption follows the water pattern, with a $T_{\rm peak}$ slightly shifted by about 1--2 K after the maximum of water desorption, and the desorption of both molecules ending mutually. This suggests that \ch{CH3CH2OH} stays bound to the water surface until the \ch{H2O} molecules leave the surface completely. The moment ethanol starts desorbing can be explained by a stronger binding energy with ASW compared to CI in the case (see below) of the lowest coverages, leading \ch{CH3CH2OH} to stay bound to water until it changes configuration by crystallizing. Please, refer to the Section \ref{ssss_exp_cry_ice} for further details about the crystallization processes.

The sub-monolayer depositions of \ch{CH3CH2NH2} (0.08, 0.13, and 0.4 ML) on the slab of 10 ML of ASW have common characteristics with the ethanol depositions. For instance, the ethylamine and \ch{H2O} desorptions end mutually, and the integrated areas of the TPD profiles of ethylamine are also proportional to the surface coverage.  However, the maximum of the desorption of \ch{CH3CH2NH2} is reached 1 K before that of water, except for the coverage of 0.08 ML, where both ethylamine and water show their maximum peak concomitantly. Additionally, in all cases, ethylamine starts desorbing before the crystallization of water at respectively 138, 140 and 132 K. Contrary to ethanol, it seems that \ch{CH3CH2NH2} binds with water in the same way, regardless of the configuration of the ice. The lowest temperature would here simply reflect the higher concentration of molecules, consequently desorbing earlier due to the unavailability of the most favourable binding sites.

However, the comparison of TPDs registered after short deposition times of both ethanol and ethylamine highlights that both molecules sublimate far later when adsorbed on water ice rather than when deposited on the gold surface, due to their more favourable and stronger interaction with the ASW ice.

\paragraph{On crystalline ice}
\label{ssss_exp_cry_ice}
$\,$
\newline
The above interpretations of a stronger binding energy of the adsorbate with ASW than with CI, in the case of low coverage for ethanol, are demonstrated experimentally by depositing 0.15 ML of \ch{CH3CH2OH} and 0.13 ML of \ch{CH3CH2NH2} on 10 ML of CI. The \ch{H2O} ice was crystallized via the process described in Section \ref{sss_water_growth}, and the results are displayed in the right panels of Figure \ref{fig:asw_ci_etoh_etnh2_0.15}. 

In the case of ethylamine, the desorption curve is identical to that on ASW. The start, end, and maximum of the desorption occur at the same temperatures, and the integrated areas are equivalent, attesting for extremely similar deposition conditions, apart from the configuration of the \ch{H2O} substrate. Thus, it seems that there is no influence of the ice structural state in the desorption energy of \ch{CH3CH2NH2} on water ice. 

In the case of ethanol, however, the \textit{m/z} = 31 signal differs fairly significantly whether deposited on ASW or on CI. On CI, although most of the sublimation into the gas phase takes place at the very end of the water desorption at $T_{\rm peak}$ = 162 K (as expected from the results obtained on ASW), a small fraction of molecules starts desorbing before water, from $\sim$117 K to $\sim$145 K. This corresponds better to the temperature range observed in the case of a deposition exclusively on gold. The curve shape suggests that when \ch{H2O} is completely crystalline, the ethanol molecules either bond slightly less (explaining their pre-desorption), or tend to agglomerate one onto each other, favouring a behaviour close to that of a multilayer regime. With any of the above solutions, the conclusion remains that in a sub-monolayer regime, ethanol has a different behaviour on ASW and on CI, further demonstrating its ejection from the surface during the phase change of water.

\subsubsection{Multilayer depositions}
\label{sss_multilayer_dep}

The desorption process of the adsorbates in a multilayer regime (i.e., 2.1 ML of \ch{CH3CH2OH} and 1.25 ML of \ch{CH3CH2NH2}) deposited onto a film of 10 ML of ASW were also performed. The corresponding TPD curves are displayed on the right panels of Figure \ref{fig:asw_etoh_etnh2_submono_multi}. The desorption profile of ethanol shows two clearly distinguishable peaks, while that of ethylamine exhibits the contribution of three desorption peaks. They all attest for different kinds of molecular interactions.

In both species, the shoulder monitored at higher temperature at the end of the \ch{H2O} desorption, furthest to the right, is comparable to the results obtained from the sub-monolayer depositions. In other words, the molecules of ethanol and ethylamine that are directly in contact with the highest energy binding sites keep being bonded very strongly with water, even when the latter is desorbing. They are therefore sublimating together with the very last water molecules.

In the case of ethylamine, the first peak observable even before the start of the water desorption corresponds to a fraction of molecules that are not directly interacting with the bulk of \ch{H2O}, but rather with the less strongly bonded multilayer. Indeed, these \ch{CH3CH2NH2}  molecules start desorbing at 102 K, corresponding well to the behaviour exhibited when a multilayer regime is deposited solely on gold, with the addition of a slight water-binding contribution, explaining the signal persistence up to $\sim$150 K (visible if using an extrapolation). The interaction between layers of \ch{CH3CH2NH2} is thus weaker than that between the adsorbate and the ASW ice film. In the case of ethanol, in contrast, this first peak characteristic of a multilayer regime is not appearing prior to the desorption of water, but is instead blended with the intermediate peak (described in the next paragraph). Indeed, on gold, multilayers of \ch{CH3CH2OH} only start to desorb around 120 K. This temperature mingles with that of the start of water desorption, therefore making the multilayer contribution indistinguishable from the second peak, when the adsorbate is deposited on \ch{H2O}.

Lastly, for both adsorbates, the largest contribution to the desorption peaks occurs at the moment of the \ch{H2O} phase change. This would correspond to the ejection of ethanol and ethylamine from the surface when water is undergoing a morphological change. In a multilayer regime, both adsorbates consequently appear to bond more strongly with ASW than with CI.

\subsubsection{Desorption energy derivation}
\label{sss_BE_derivation}


The desorption conditions on water ice do not allow to easily extract the desorption energies, E$_{\rm des}$, of the adsorbates. Indeed, in most of the cases, the interaction between the adsorbates and water ice is stronger than that of water with itself, resulting in a co-desorption of \ch{H2O} and the adsorbates, or in a late desorption of the adsorbates, together with the last \ch{H2O} molecules present on the sample. Since water starts sublimating first, calculating the E$_{\rm des}$ of the adsorbates on water ice is therefore unfeasible \cite{minissale2022}. Likewise, the multiple peaks found when several adsorbate layers are deposited also prevent from a clear derivation of the adsorbate/water E$_{\rm des}$.

However, obtaining E$_{\rm des}$ values is possible by using the sub-monolayer and monolayer depositions of \ch{CH3CH2OH} and \ch{CH3CH2NH2} on the bare gold substrate (see above). 
The mathematical model, based on Eqn. \ref{<Polanyi_Eq>} and developed by \citet{chaabouni2018}, was employed to fit the experimental results. The code returns different populations (N) of molecules sublimating into the gas phase with a certain desorption energy (E$_{\rm des}$), all together contributing to the final TPD curve. 
The best fit of the experimental data is obtained when the calculated curve matches well the experimental results, which is satisfied when the three parameters ($A_{\rm TST}$; E$_{\rm des}$, N) are well constrained.

The pre-exponential factors needed to obtain the curves and the populations were calculated with an equation derived from the transition state theory \cite{tait2005} (further details can be found in the ESI$^\dag$), yielding $A_{\rm TST}$ = 3.31 $\times$ 10$^{18}$ s$^{-1}$ for ethanol, and $A_{\rm TST}$ = 2.04 $\times$ 10$^{18}$ s$^{-1}$ for ethylamine.

\begin{figure*}[ht]
\centering
  \includegraphics[width=\textwidth]{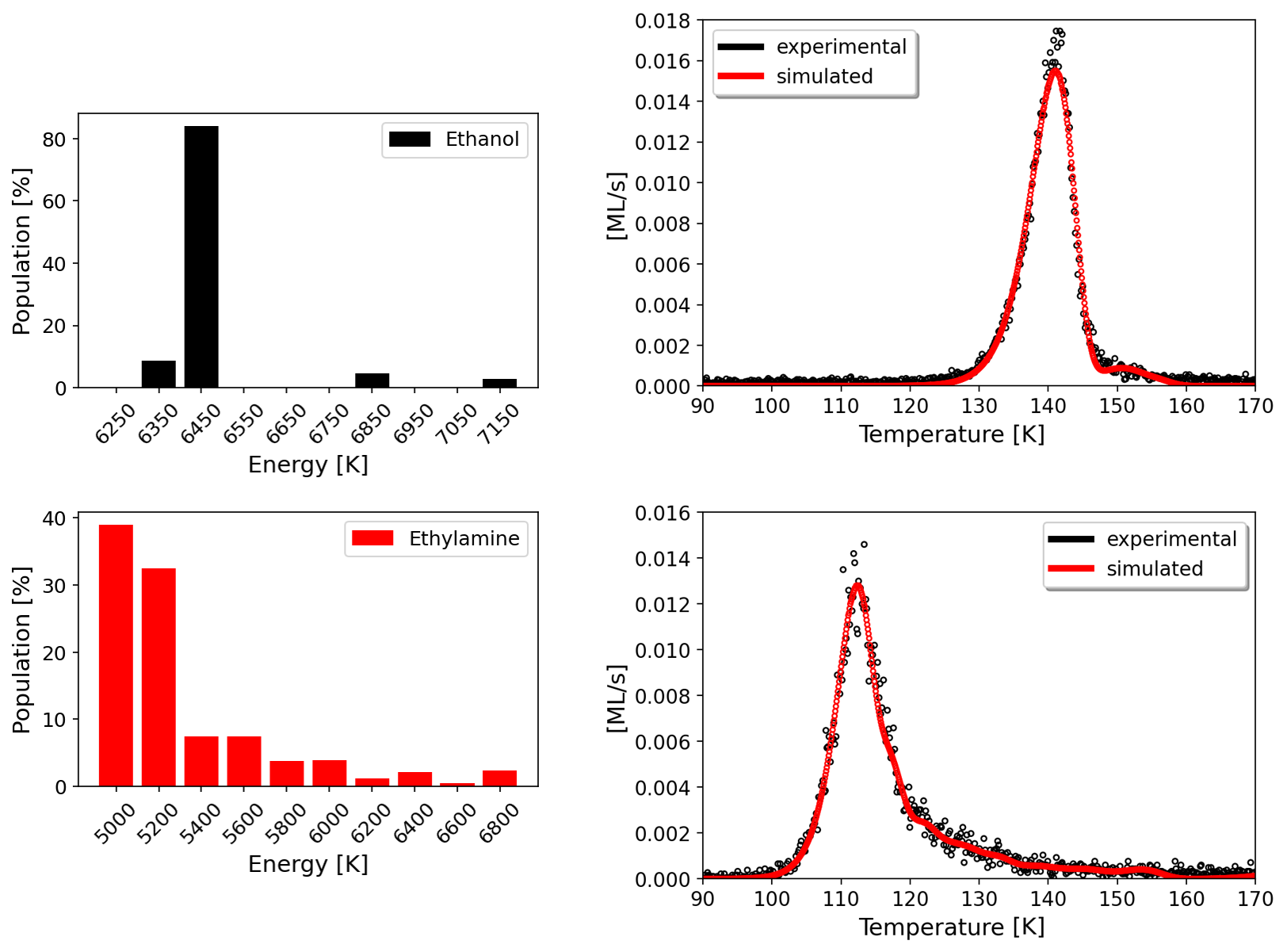} \\
  \caption{Comparison between the measured and reproduced TPD profiles and relative populations of 0.7 ML of \ch{CH3CH2OH} adsorbed on gold (top panel, 8 TPDs) and 0.4 ML of \ch{CH3CH2NH2} adsorbed on gold (bottom panel, 10 TPDs), considering the calculated pre-exponential factor.}
  \label{fig:python_code}
\end{figure*}

The model returns E$_{\rm des}$ =  53.6 kJ mol$^{-1}$ (6450 K) for \ch{CH3CH2OH} and E$_{\rm des}$ = 41.6--43.2 kJ mol$^{-1}$ (5000--5200 K) for \ch{CH3CH2NH2}, as displayed in Figure \ref{fig:python_code} (right panels), where the best fits of the simulated TPD (in red) match well the ones measured in the experiments (represented by the black dots). Each simulated curve is given by the combination of a number of TPDs, each corresponding to the different E$_{\rm des}$ obtained in the population graph (left panel of Figure \ref{fig:python_code}). For ethanol, we do not obtain an actual distribution, as more than 80\% of the molecules are characterized by one main energy value. On the other hand, ethylamine shows a proper distribution with almost 70\% of the population having E$_{\rm des}$ between 41.6--43.2 kJ mol$^{-1}$ (5000--5200 K), followed by about 15\% of the population with E$_{\rm des}$ between 44.9--46.6 kJ mol$^{-1}$ (5400--5600 K) and a queue in which each E$_{\rm des}$ has a progressively smaller weight until reaching a negligible contribution.

\section{Discussion}\label{sec:discussion}
\begin{table*}[htb]
\centering
\small
  \caption{Summary of all the computed quantities. All quantities are ZPE corrected and are provided in kJ mol$^{-1}$ and in Kelvin. BE(0) = binding energy. E$_C$ = Cohesive Energy; E$_X$ = Extraction Energy; ML = monolayer; CI = crystalline ice; ASW = amorphous solid water. S-ADS = single adsorption; H-ML = half monolayer; F-ML = full monolayer. The values that constitute the endpoints of a range are separated by a dash, while multiple values are separated by a slash.}
  \label{tab:summary}
  \begin{tabular*}{\textwidth}{@{\extracolsep{\fill}}lllll|llll}
    \hline
    & \multicolumn{4}{c}{ kJ mol$^{-1}$} & \multicolumn{4}{c}{Kelvin} \\
    & \ch{CH3CH2OH} & \ch{CH3CH2NH2} & CI & ASW & \ch{CH3CH2OH} & \ch{CH3CH2NH2} & CI & ASW\\
    \hline
    Bulk E$_C$ & -53.4 & -44.0 & -55.8 & &   -6423 & -5292 & -6711 & \\
    Surface E$_C$ & -49.4 -- -40.3 & -40.4 -- -30.3 &  -49.5 &  -44.9 &   -5941 -- -4847 & -4859 -- -3644 & -5953 & -5400 \\
    Surface E$_X$ & 58.9 -- 116.6 &  36.0 -- 84.3 & &  &  7084 -- 14024 &  4330 -- 10139 & & \\
   \hline
    BE(0) on (010) \ch{EtOH} &  20.9 / 30.9 / 41.8 &  & & &  2514 / 3716 / 5027 & & & \\
    BE(0) on (100) \ch{EtNH2} & &  21.4 / 21.6 / 43.0 &  & & & 2574 / 2598 / 5172 & & \\
   \hline
    BE(0) S-ADS on CI &  33.4 / 42.2 / 61.6 &   23.1 / 61.4 & & &  4017 / 5075 / 7409  & 2778 / 7385 & & \\
    BE(0) H-ML on CI & 66.0 &  64.5 &  & & 7938 & 7758 &  & \\
    BE(0) F-ML on CI & 50.0 &  49.5 &  & & 6014 & 5953 & & \\
    BE(0) S-ADS on ASW & 26.0 -- 59.1 &  19.1 -- 71.7 &  & & 3127 -- 7108 & 2297 -- 8624 & & \\
    \hline
  \end{tabular*}
\end{table*}

\subsection{Pure \ch{CH3CH2OH} and \ch{CH3CH2NH2} depositions}

The desorption of \ch{CH3CH2OH} and \ch{CH3CH2NH2} deposited on the bare gold surface resulted in a TPD with a $T_{\rm peak}$ of 141 K and 113 K, respectively. From there,  E$_{\rm des}$ = 53.6 kJ mol$^{-1}$ for ethanol and E$_{\rm des}$ = 41.6--43.2 kJ mol$^{-1}$ for ethylamine were derived, along with the respective pre-exponential factors $\nu$. From a computational point of view, this scenario can be reproduced in two ways: i) by modeling the bulk and the surfaces of EtOH and EtNH$_2$ crystals; or ii) by adsorbing \ch{CH3CH2OH} and \ch{CH3CH2NH2} on the most stable EtOH and EtNH$_2$ surfaces. The quantity that determines the strength of the interaction between the molecules forming the crystals is the cohesive energy E$_C$ of the bulk (-53.4 kJ mol$^{-1}$ for EtOH and -44.0 kJ mol$^{-1}$ for EtNH$_2$) and of the most stable surfaces (-49.4 kJ mol$^{-1}$ for EtOH and -40.4 kJ mol$^{-1}$ for EtNH$_2$). It should be noted that the negative values taken by E$_C$ are due to its definition, which is the opposite of that of the BE(0). Therefore, a more negative value indicates a more stable structure. We also simulated the adsorption of \ch{CH3CH2OH} on the (010) EtOH surface, and \ch{CH3CH2NH2} on the (100) EtNH$_2$ surface, to compute the relative binding energy, which resulted in a range covering similar values for the two species (20.9--41.8 kJ mol$^{-1}$ for \ch{CH3CH2OH} and 21.4--43.0 kJ mol$^{-1}$ for \ch{CH3CH2NH2}). Despite the similarity of the BE(0)s, the first calculations suggest that the interaction between organized ethanol molecules is stronger than that between ethylamine molecules. 
This is in agreement with the E$_{\rm des}$ of ethanol and ethylamine estimated experimentally. Remarkably, the E$_C$ of the bulk and of the most stable EtOH and EtNH$_2$ surfaces seems to be the quantity that better describes what we observe in the experiment, supposing that small quantities of adsorbate, when deposited on the gold finger, or during the heating phase, could aggregate and order themselves into structures that resemble the order present in the crystal. 


\subsection{Sub-monolayer depositions}

The experiments performed in this work cannot provide an experimental measurement of the E$_{\rm des}$ of \ch{CH3CH2OH} and \ch{CH3CH2NH2} adsorbed on water ice. This is because the interpretation of the TPD profiles is hindered by the presence of co-desorption phenomena, and occasionally the emergence of multiple desorption peaks.
This occurred as well in previous experimental studies (e.g., \citet{lattelais2011}) investigating the desorption of ethanol from water ice. For this reason \citet{burke2015} classified ethanol as a complex water-like molecule, in that it wets the ice surface and co-desorbs with water (which moreover affects the water crystallization process), like methanol, acetic acid, and other iCOMs. In \citet{minissale2022} is also mentioned that the species capable of establishing several H-bonds with water are likely to strongly bound with the ice and behave like ethanol, thereby making impracticable the determination of their E$_{\rm des}$ from a TPD experiment. From our calculations, this phenomenon can be explained by the presence of BE(0) values of \ch{CH3CH2OH} and \ch{CH3CH2NH2} responsible for an interaction between the adsorbate and water that is stronger than the cohesive energy of water ice (-55.8 kJ mol$^{-1}$ in the case of CI bulk, and slightly inferior for ASW and CI surfaces).

Moreover, the experiments show that \ch{CH3CH2NH2} binds with ASW and CI with similar strength, regardless of their different structural state. On the contrary, \ch{CH3CH2OH} seems to have stronger binding energies with ASW compared to CI in the case of low coverages (0.15 and 0.25 ML). This aspect does not clearly emerge from the calculations, because the differences in the BE(0) ensembles of ASW and CI are limited. While the BE(0) values computed for ethylamine cover almost the same range for ASW (19.1--71.7 kJ mol$^{-1}$) and CI (23.1 / 61.4 kJ mol$^{-1}$), ethanol can experience lower BE(0)s on ASW (26.0--59.1 kJ mol$^{-1}$) rather than on CI (33.4 / 42.2 / 61.6 kJ mol$^{-1}$), thus in disagreement with the experimental data.

Additionally, an aspect that must be considered in the comparison of experimental and theoretical results is that, in a sub-monolayer deposition TPD experiment, constant thermal energy is provided to the molecules, which is used to diffuse and move toward stronger binding sites, where the system is more stable, this way neglecting the least favourable binding sites. Moreover, the experiments measure desorption energies, accounting for desorption effects that cannot be reproduced in the calculations. Therefore, the E$_{\rm des}$ provided by the experiments do not usually cover the lower limit of the BE(0) distribution obtained with calculations.

In this sense, quantum chemical calculations offer the possibility to compute the quantities that cannot be determined experimentally. 
Despite their structural similarity, \ch{CH3CH2OH} and \ch{CH3CH2NH2} show differences in their chemical behaviour, due to the H-bond capabilities of the OH and NH$_2$ groups. NH$_2$ is a stronger H-bond acceptor than OH, but it is a weaker H-bond donor. This results in \ch{CH3CH2NH2} spanning a larger BE(0)s range than \ch{CH3CH2OH} when adsorbed on the ASW model (19.1--71.7 kJ mol$^{-1}$ vs 26.0--59.1 kJ mol$^{-1}$, respectively) since the strength of the interaction depends on the atom establishing the H-bond with the binding site of the ice. 
Such difference can also partially be observed with the CI model, where the lowest BE(0), due to a H-bond donation from the adsorbate to the surface, is 23.1 kJ mol$^{-1}$ for ethylamine and 33.4 kJ mol$^{-1}$ for ethanol. The values obtained for the CI can represent in part the BE ensemble obtained on the ASW, although two or three values are not sufficient to account for the diversity of binding sites available on the amorphous model.

The BE(0)s obtained for \ch{CH3CH2OH} and \ch{CH3CH2NH2} on CI and ASW mostly agree with those proposed in the literature. \citet{garrod2013} suggested BE = 65.5 kJ mol$^{-1}$ for \ch{CH3CH2NH2} and BE = 52.0 kJ mol$^{-1}$ for \ch{CH3CH2OH}, highlighting the stronger interaction of ethylamine, compared to ethanol, with the ice. To the best of our knowledge, no other BEs have been documented for \ch{CH3CH2NH2} on water ice. At variance, ethanol has been the object of a number of investigations. \citet{lattelais2011} computed the adsorption of ethanol onto the same crystalline proton-ordered ice model of this work, using the PW91 functional in combination with plane waves, with the VASP code, obtaining BE = 56.5 kJ mol$^{-1}$, close to the value of \citet{garrod2013}. Due to the use of plane waves, the BE computed by \citet{lattelais2011} 
does not suffer from the BSSE. However, at variance with our calculations, both the ZPE and the dispersion corrections are lacking, probably explaining the discrepancy with the value computed in this work for the same adsorption complex (BE(0) = 61.6 kJ mol$^{-1}$).

On the contrary, the values provided by \citet{etim2018}, in which ethanol is interacting with only one water molecule, range from 37.4 to 40.7 kJ mol$^{-1}$, which lay within the intermediate range of our BE(0)s ensemble. Clearly, the latter estimate cannot reproduce the largest BE(0)s, due to the impossibility of capturing the essence of the interaction between \ch{CH3CH2OH} and water ice when approximating its structure with only one water molecule. The choice of the ice model appears to be crucial, especially when the adsorbate is large enough to interact with more than one water molecule. Thus, the values proposed in this work represent an extension of those already available in the literature, additionally describing a tail of weaker BE(0) values that can only be simulated when using extended amorphous ice models characterized by a rich structural variability.

The simulation of the H-ML and F-ML scenarios by adsorbing a larger number of molecules on the surface yields slightly larger and lower BE(0)s compared to the S-ADS case, respectively. In the calculations, we directly modeled the adsorption of a layer of molecules, which results in the determination of a single BE(0) value for all the species adsorbed. This does not faithfully represent what happens in the experiment, where the species approach the surface in different moments and start occupying binding sites starting from the strongest and then to the weakest, resulting in a distribution of BEs. However, the BE(0)s obtained from the simulation of different surface coverages represents a useful alternative to understand the variation in the BE(0) values resulting from the occupation of an increasing number of surface binding sites.

\subsection{Multilayer depositions}  

The deposition of 1.25 ML of \ch{CH3CH2NH2} on ASW results in multiple desorption peaks, in which the desorption of the adsorbate alone can be discerned from its co-desorption with water. The first desorption peak (starting at 102 K, before water desorption) corresponds to ethylamine molecules that are not in direct contact with the water surface. The interaction between layers of \ch{CH3CH2NH2} is thus weaker than that between \ch{CH3CH2NH2} and the ASW film. However, in the TPD of 2.1 ML of \ch{CH3CH2OH} on ASW, the peak corresponding to the multilayer desorption is blended with that caused by the desorption of the ethanol in direct contact with water ice due to crystallization of the latter. This causes the two types of desorption to be indistinguishable. Such discrepancy between \ch{CH3CH2NH2} and \ch{CH3CH2OH} can be justified by comparing, once again, the E$_C$ of the species under investigation. The computed cohesive energies of ASW and CI surfaces (-49.5 and -44.9 kJ mol$^{-1}$, respectively) are comparable with those of ethanol surfaces (between -49.4 and -40.3 kJ mol$^{-1}$), but larger than those of ethylamine surfaces (between -40.4 and -30.3 kJ mol$^{-1}$). Based on these numbers, we can assume that even in the presence of larger quantities of ethanol on the water ice surface, both ethanol and water likely desorb in the same temperature range: the multilayer desorption is regulated by the cohesive energy, while the desorption of those ethanol molecules in direct contact with water is determined by the BE(0) of \ch{CH3CH2OH} on ASW.

\subsection{Astrophysical Implications}

The experiments performed in this work reveal that ethanol and, in part, ethylamine co-desorb with water ice. Computational data confirm and complement the experimental results, providing an ensemble of BE(0)s accounting for the upper and lower limits. Usually, the weakest interactions between the adsorbate and the surface cannot be measured by TPD experiments, but needs to be considered when determining the regions of the ISM in which a species is available in the gas phase, rather than segregated on the ice grains. Despite this, the presence of ethanol, as well as other iCOMs in general, in the cold (<20 K) outskirts of prestellar cores requires non-thermal desorption processes to be explained. 

In contrast, ethylamine has only been tentatively detected towards the Galactic center cloud G+0.693-0.027,\cite{zeng2021} raising doubts regarding its absence elsewhere. This work suggests that both \ch{CH3CH2OH} and \ch{CH3CH2NH2} should be found in warm regions, like hot cores and hot corinos, where the ice mantles sublimate due to the presence of a warm central object. 

The E$_{\rm des}$ of the species on water ice mantles have a profound impact on the chemical composition of the regions where planetary systems eventually form. A very important case is represented by the gaseous-versus-solid chemical composition of protoplanetary disks. If a species is in the solid form, it will likely be incorporated in rocky planets, asteroids and comets, whereas, if it resides in the gas phase, it will only enrich the giant gaseous planets. The transition of a species from the gas phase to the solid phase (when moving away from the central object) is identified by its snow-line.\cite{oberg2021} 

If we compare the BE(0) ranges spanned by ethanol (26.0--59.1 kJ mol$^{-1}$), ethylamine (19.1--71.7 kJ mol$^{-1}$) and water (14.2--61.6 kJ mol$^{-1}$, peaking at 35.2 kJ mol$^{-1}$),\cite{tinacci2023} we can have clues on the composition of the icy mantles on the protoplanetary disks, dictating whether a species is incorporated in water-rich planetesimals or not. By looking at the maximum BE(0) value of \ch{CH3CH2OH} and \ch{CH3CH2NH2}, we can infer that at least a fraction of these species will remain on the grains in regions where up to 96\% of water has completely sublimated (corresponding to BE(0) $\simeq$ 54.0 kJ mol$^{-1}$). Ethanol and ethylamine could therefore be incorporated in the planetesimals that will form rocky bodies.

Theoretical calculations indicate that the BE(0) ensembles of \ch{CH3CH2OH} and \ch{CH3CH2NH2} on water ice span similar ranges. Accordingly, if ethanol is detected in the ice,\cite{mcclure2023} also ethylamine should be detected. However, this is not the case. Thus, the reason behind the presence of ethanol and the absence of ethylamine in ISM regions point towards their synthetic pathways. As mentioned in Section \ref{sec:intro}, ethanol synthesis on water ices has been proposed via reactivity between CCH and \ch{H2O} followed by hydrogenation.\cite{perrero2022ethanol} The analogue reaction for ethylamine would be thus initiated by CCH + NH$_3$. However, ammonia is much less abundant than water, causing a meager production of \ch{CH3CH2NH2}. Hydrogenation of icy acetaldehyde (CH$_3$CHO), which can form by the HCO + CH$_3$ radical-radical coupling on water ice,\cite{enrique-romero2021} is an alternative route towards ethanol formation. The analogue pathway for ethylamine is the hydrogenation of acetonitrile (CH$_3$CN), which can form previously through HCN + CH$_3$ or CN + CH$_3$ reactions. However, HCN is chemically inert in cold environments,\cite{smith2001} while the likelihood of CH$_3$CN undergoing hydrogenation is doubtful, due to the intrinsic stability (and accordingly inertness) of nitriles.\cite{nguyen2019} Moreover, detection of CH$_3$CN in ices still remains elusive.\cite{mcclure2023} Additionally, at variance with ethanol, ethylamine could form salts due to the basicity of the amino moiety. In general, O-bearing species are more abundant than N-bearing species, a fact directly linked to the cosmic abundances of oxygen and nitrogen,\cite{cardelli1993,meyer1998,jenkins2009} although such discrepancy cannot fully account for the differences in the detection of these two iCOMs, especially considering the extremely articulated network of chemical reactions taking place in the ISM.

\section{Conclusions} \label{sec:conclusions}

In this work, the interaction of \ch{CH3CH2OH} and \ch{CH3CH2NH2} on water ice has been studied by means of a computational (periodic DFT simulations) and an experimental (TPD measurements) approach.

First, the E$_{\rm des}$ of pure \ch{CH3CH2OH} and \ch{CH3CH2NH2} deposited on a gold surface were determined experimentally, which closely align well with the cohesive energies of crystalline bulks of EtOH and EtNH$_2$, along with their most stable surfaces.

The E$_{\rm des}$ of submonolayer coverages of \ch{CH3CH2OH} and \ch{CH3CH2NH2} deposited on crystalline and amorphous water ice surfaces cannot be determined experimentally due to the complexity of their desorption profiles, where multiple desorption peaks are occasionally present, in addition to the co-desorption of the adsorbate with water. 
Theoretical calculations fill this gap by providing a comprehensive set of BE(0) values for \ch{CH3CH2OH} and \ch{CH3CH2NH2} on both crystalline and amorphous surface ice models. Notably, these calculations reveal weak binding sites that are elusive in traditional TPD experiments. Nevertheless, the TPD of a submonolayer deposition of ethanol on crystalline ice reveals a very small fraction of molecules desorbing from the surface prior to its crystallization, indicating a weaker interaction of ethanol with crystalline ice compared to the amorphous one. 
The deposition of a multilayer of adsorbate can provide additional information, due to the occupation of all the binding sites available on the water surface and the formation of multiple layers of adsorbate, avoiding a direct interaction with the water ice. The temperature at which the first ethylamine molecules of the multilayer desorb is higher than that of pure ethylamine, owing to the presence of the water ice surface. After desorption of the multilayer, the TPD profile resembles that of a submonolayer deposition. Conversely, for ethanol, it is not possible to distinguish the two behaviours as the corresponding peaks are overlapping.
The theoretical results corroborate these findings, in that the cohesive energy of bulk and surfaces of crystalline EtNH$_2$ is smaller compared to that of EtOH, which in turn is comparable with that of crystalline and amorphous water ice. This disparity explains why ethylamine multilayers desorb before water, while ethanol multilayers do not.

The desorption energies of \ch{CH3CH2OH} and \ch{CH3CH2NH2} play a critical role in determining their presence in the gas or in the solid phase of the ISM, thus influencing the chemical composition of celestial bodies formed from protoplanetary disks, as delineated by snow-lines. The high BE(0)s of ethanol rationalize its presence in interstellar ices. However, the non-detection of ethylamine cannot be explained by its BE(0)s, and likely stems from the lack of favourable reaction pathways responsible for its formation. Nevertheless, the scarcity of gas-phase \ch{CH3CH2NH2} in the ISM and the presence of \ch{CH3CH2OH} in the gas phase of cold objects, the latter usually elucidated by invoking non-thermal desorption mechanisms, needs further investigation. 

This work exemplifies the way in which not only computational chemistry can support laboratory experiments, but the two disciplines can complement each other in order to provide reference data as input for astrochemical models.\cite{fortenberry2024QMmatch}

\section*{Author Contributions}
Conceptualization: J. P., J. V., A. R., and F. D.; experiments: J. P., J. V., and E. C.; data
curation: J. P., J. V., A. R., and F. D.; formal analysis: J. P., J. V., A. R., P. U., and F. D.; funding acquisition: F. D.; investigation: J. P., J. V., A. R., P. U., and F. D.; methodology: J. P., J. V., A. R., and F. D.;
validation: A. R., P. U., E. C. and F. D.; writing – original draft: J. P., and J. V.

\section*{Conflicts of interest}
There are no conflicts to declare.

\section*{Acknowledgements}
This project has received funding within the European Union’s Horizon 2020 research and innovation programme from the European Research Council (ERC) for the project ``Quantum Chemistry on Interstellar Grains” (QUANTUMGRAIN), grant agreement No 865657. This research is also funded by the Spanish MICINN (projects PID2021-126427NB-I00 and CNS2023-144902). The Italian Space Agency for co-funding the Life in Space Project (ASI N. 2019-3-U.O), the Italian MUR (PRIN 2020, Astrochemistry beyond the second period elements, Prot. 2020AFB3FX) are also acknowledged for financial support. Authors (J.P. and P.U.) acknowledge support from the Project CH4.0 under the MUR program ``Dipartimenti di Eccellenza 2023-2027” (CUP: D13C22003520001). This work was also funded by CY Initiative of Excellence (grant "Investissements d'Avenir" ANR-16-IDEX-0008), Agence Nationale de la recherche (ANR) SIRC project (Grant ANR-SPV202448 2020-2024), by the Programme National "Planétologie" and "Physique et Chimie du Milieu Interstellaire" (PCMI) of CNRS/INSU with INC/INP co-funded by CEA and CNES.
The RES resources provided by BSC in MareNostrum to activities QHS-2022-3-0007 and QHS-2023-2-0011, and the supercomputational facilities provided by CSUC and CINECA (ISCRAB projects) are also acknowledged. The EuroHPC Joint Undertaking through the Regular Access call project no. 2023R01-112, hosted by the Ministry of Education, Youth and Sports of the Czech Republic through the e-INFRA CZ (ID: 90254) is also acknowledged.
J.P. and J.V. wish to thank Michel Lorin and Emilie Schmitter for their significant contribution in performing the experiments. 



\balance


\bibliography{main} 
\bibliographystyle{rsc} 

\providecommand*{\mcitethebibliography}{\thebibliography}
\csname @ifundefined\endcsname{endmcitethebibliography}
{\let\endmcitethebibliography\endthebibliography}{}

\end{document}